\documentclass[traditabstract]{aa}
\usepackage{natbib}
\bibpunct{(}{)}{;}{a}{}{,}%
\usepackage{graphicx}
\usepackage{txfonts,color}
\usepackage{subfigure}
\usepackage{multirow}
\begin{document}
\title{Acoustic power absorption and enhancement generated by slow and fast MHD waves.}
\subtitle{Evidence of solar cycle velocity/intensity amplitude changes consistent with the mode conversion theory}
\author {R.~Simoniello\inst{1}
\and W.~Finsterle\inst{1}
\and R.~A.~Garc\'ia\inst{2,3}
\and D.~Salabert\inst{4,5}
\and A.~Jim\'enez\inst{4,5}
\and Y.~Elsworth\inst{6}
\and H.~Schunker\inst{7}}
\institute{PMOD/WRC Physikalisch-Meteorologisches Observatorium Davos-World Radiation Center, 7260 Davos Dorf, Switzerland \\
\email rosaria.simoniello@pmodwrc.ch, \email wolfgang.finsterle@pmodwrc.ch
\and Laboratoire AIM, CEA/DSM-CNRS-Universit\'e Paris Diderot; CEA, IRFU, SAp, centre de Saclay, F-91191, Gif-sur-Yvette, France
\and GEPI, Observatoire de Paris, CNRS, Universit\'e Paris Diderot; 5place Jules Janssen, 92190 Meudon, France\\
\email rgarcia@cea.fr
\and IAC, Instituto de Astrofis\'\i ca de Canarias, 38205, La Laguna, Tenerife, Spain
\and Departamento de Astrof\'isica, Universidad de La Laguna, E-38205 La Laguna, Tenerife, Spain\\
\email salabert@iac.es, ajm@iac.es
\and School of Physics and Astronomy, University of Birmingham, Edgbaston, Birmingham B15 2TT, UK \\
\email ype@bison.ph.bham.ac.uk
\and Max-Planck-Institute for Solar System Research, Max-Planck-Strasse 2, Katlenburg-Lindau, 37197, Germany\\
\email schunker@mps.mpg.de}
\date{Received, accepted}
\abstract
{We used long duration, high quality, unresolved (Sun-as-a star) observations collected by the ground based network BiSON and by the instruments GOLF and VIRGO on board the ESA/NASA SOHO satellite  to search for solar-cycle-related changes in mode characteristics in velocity and continuum intensity for the frequency range between 2.5~mHz $<\nu<$ 6.8~mHz. Over the ascending phase of solar cycle 23 we found a suppression in the $p$-mode amplitudes both in the velocity and intensity data between 2.5~mHz $<\nu<$ 4.5~mHz with a maximum suppression for frequencies in the range between 2.5~mHz $<\nu<$ 3.5~mHz. The size of the amplitude suppression is 13$\pm 2$ per cent for the velocity and 9$\pm 2$ per cent for the intensity observations. Over the range 4.5~mHz $<\nu<$ 5.5~mHz the findings hint within the errors to a null change both in the velocity and intensity amplitudes. At still higher
frequencies, in the so called {\bf H}igh-frequency {\bf I}nterference {\bf P}eak{\bf s} (HIPs) between 5.8~mHz $<\nu <$ 6.8~ mHz, we found an enhancement in the  velocity amplitudes with the maximum 36$\pm 7$ per cent occurring for 6.3~mHz $<\nu<$ 6.8~mHz. However, in intensity observations we found a rather smaller enhancement of about 5$\pm$2 per cent in the same interval. There is evidence that the frequency dependence of solar-cycle velocity amplitude changes is consistent with the theory behind the mode conversion of acoustic waves in a non-vertical magnetic field, but there are some problems with the intensity data, which may be due to the height in the solar atmosphere at which the VIRGO data are taken.}
\keywords{Sun: oscillations - Sun: surface magnetism - methods: observational - methods: data analysis}
\titlerunning{Velocity and intensity amplitude changes}
\authorrunning{Simoniello et al.}
\maketitle
\section{Introduction}
The Sun shows a rich spectrum of solar oscillation modes usually referred to as $p$-modes \citep{Lei71,Ulr70,Lei62,Deu75}. They are  stochastically excited and intrinsically damped by turbulent convection \citep{Gol77} resulting in a significant variation in their power with time.  In addition, however, there is some evidence that highly energetic events can excite both high-degree modes \citep{Kos98} and even global modes \citep{Fog98,Kar07}.

It was predicted that only those waves with frequencies below the acoustic cut-off frequency \citep[measured in the Sun as $\nu_{ac}\approx$ 5.3~mHz, e.g.][]{Jima06}  would be trapped inside the acoustic cavity of the Sun \citep{Bal90}; while  higher frequency waves, above $\nu_{ac}$, could travel freely in the solar atmosphere. Hence the latter became a possible  source for  chromospheric heating \citep{Alf47,Sch48,Ulm90}.
Simulations show that photospheric oscillations can produce chromospheric shocks through upward propagation \citep{Car97}, but the role this plays in chromospheric heating is still under debate \citep{Fos05,Kal07}.

Through observations and with the help of analyses a variety of types 
of oscillation modes with frequencies $\nu<\nu_{ac}$ were recently detected. It was claimed
that these are slow-magnetoacoustic waves generated by mode transmission \citep{Cal95,Cal08} propagating in the
atmosphere, because the inclined magnetic field supposedly reduced
the effective acoustic cut-off \citep{Dep04,Jef06}. The transfer of energy between internal 
acoustic
and external slow magnetoacoustic waves is a function of frequency,   
thickness of the interaction region (where the Alfv\'en speed is  
comparable
to the sound speed) and
angle between the wave vector and the magnetic field vector
(attack angle).  A narrow attack angle, lower frequency and thin  
interaction region yields enhanced transmission from internal  
acoustic waves to slow external magnetoacoustic waves \citep{sch06}.

This mechanism is the preferred one to explain the lack of acoustic power observed in sunspots compared to the quiet Sun between 2~mHz $<\nu<$ 5~mHz both in intensity and velocity power maps \citep{Woo81,Lit82,Bro92}. Additionally, an acoustic enhancement above 5~mHz in the velocity amplitudes has been found in and/or around sunspots and active regions, mainly in pixels with intermediate and weak magnetic field strengths. Nothing similar was found in the continuum intensity \citep{Bra92,Hin98,Tho00,Jai02}.

Several questions are still open: Are there any localized sources that enhance the acoustic emissivity? Is the same mechanism behind the acoustic absorption and enhancement? Why is there a lack of enhancement in continuum intensity data? Are the waves incompressible? What, if any,  is the influence on the data of the height in the solar atmosphere at which the observations are localized?
In this paper we attempt to address these issues by investigating the induced  solar cycle changes in velocity  and intensity mode amplitudes for frequencies between 2.5~mHz $<\nu<$ 6.8~mHz. We do not limit ourselves to a study of the phenomena in sunspots. Instead we decided to use integrated sunlight observations, because, although the integration over the solar surface will be dominated by the quiet Sun, \citep{Unn59,ste82,Alm04} mode conversion could take place because of the network and strong localized magnetic fields with strengths of up to 2~kG \citep{Alm04}. The simultaneous observation of both absorption and enhancement could be a hint that the same physical mechanism is behind them. Equally, if we get evidence that at any one time only one of the two phenomena is observed, this would support the theory of differing mechanisms for two processes.

Furthermore integrated sunlight observations are dominated by low-$\ell$ modes.  The long duration, high quality data available in integrated sunlight measurements have already demonstrated the interaction between acoustic waves and the solar magnetic activity cycle \citep{Els93,Cha00,Kom00,Jim03,Sal06}. These modes are essentially vertical at the surface and so the inclination of the field at the surface is equal to the attack angle.

An additional real novelty of this investigation is the characterization of solar cycle changes in the velocity and intensity mode amplitudes above the acoustic cut-off, and for the first time  the results from integrated sunlight data are compared with the theoretical predictions of mode conversion. Our findings point to acoustic suppression in the $p$-band (2.5~mHz $<\nu<$ 4.5~mHz), and acoustic enhancement in the high frequency band above 5.8~mHz.
\section{Integrated sunlight observations}
\subsection{Ground-based network}
The {\bf Bi}rmingham {\bf S}olar {\bf O}scillation {\bf N}etwork (BiSON) has been operating since 1976 and makes unresolved solar disk observations. It consists of six resonant scattering spectrometers, which perform Doppler velocity measurements in integrated sunlight on the K Fraunhofer line at 7699\AA~\citep{Cha95}. The data are dominated by the Doppler variations from the low-degree $\ell$ modes.
The nominal height in the solar atmosphere at which the measurement is made is  $\approx$260~km above the photosphere \citep[e.g.][]{Jim07} as defined by  $\tau=1$ at 500nm . The data provided by the BiSON network for this investigation cover the period from April 1996 until July 2006.
\subsection{Satellite instruments}
The {\bf G}lobal {\bf O}scillation at {\bf L}ow {\bf F}requency (GOLF) instrument on board the {\bf SO}lar and {\bf H}eliospheric {\bf O}bservatory (SOHO) satellite is devoted to the search of low-degree modes. It works by measuring the Doppler shifts of the Na line at 5889\AA (D1) and 5896\AA (D2) \citep{Gab95}. Due to the malfunctioning of the polarization system, GOLF is working in single-wing configuration, which has  changed during the  10 years of observations as follows: 1) 1996-1998 in the blue-wing configuration; 2) 1998-2002 in the red-wing configuration; 3) 2002 until today in the blue-wing configuration \citep{Ulr00,Gar05}. Due to the formation heights of the spectral lines GOLF observes at $\approx$330~km in the blue-wing configuration, but at $\approx$ 480~km in the red-wing configuration \citep{Jim07}. For this survey we will use only the first blue-wing period together with the red-wing. The data provided by the GOLF instrument for this investigation cover the period from April 1996 until July 2002. The optimum calibration of the second blue-wing period is still under investigation.

The {\bf V}ariability of Solar {\bf IR}radiance and {\bf G}ravity {\bf O}scillation (VIRGO) instrument is ma\-de of three {\bf S}un {\bf P}hoto{\bf M}eters (SPM) at 402nm (blue), 500nm (gre\-en) and 862nm (red) that look at the Sun as a star \citep{Fro95}. In this paper we refer to the three VIRGO channels as VIRGO(red), VIRGO(green) and VIRGO(blue).
Ea\-ch one of the three different signals observes the Sun-as-a-star oscillations at different heights in the solar photosphere \citep{Jima05}. Response functions (RFs) are a powerful tool for the analysis of the information content and diagnostic potential of spectral lines. These functions measure the reaction of the line profile when the atmosphere is locally perturbed at a given height \citep{Soc98}. Figure 1 shows the RFs for the three VIRGO channels in their respective colors. Although the RFs are broad, the observational height can be approximated to be at the maximum value of the RFs for the three channels. Therefore we can state that the blue channel observes at $\approx$-26km, the green channel at $\approx$-10km and the red channel at $\approx$+11km.
Furthermore the blue and green channels are more sensitive to the lower part of the solar photosphere, whereas the red channel is more sensitive to slightly higher photospheric layers.
The data provided by the VIRGO observations cover the full solar cycle 23. In this survey we use the data provided by all three VIRGO channels. The data provided by VIRGO for this investigation cover the period starting from April 1996 until July 2006.
\begin{figure}
\centering
\includegraphics[width=3in]{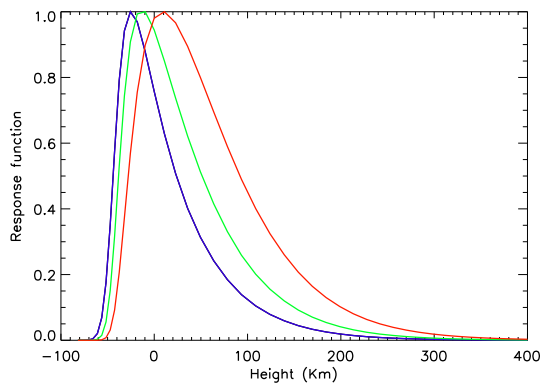}
\caption{The three response functions for the three VIRGO channels (blue, green, red) indicated by their respective colors.} \label{fig:rf}
\end{figure}
\section{Data Analysis}
\subsection{Mean Fourier transform at low frequency}
We are interested in investigating the frequency dependence of the velocity and intensity amplitude  with the solar activity cycle. Therefore we split the entire frequency range into five intervals shown in Table 1.
\begin{table}
\begin{center}
\caption{The frequency bands used for the data analysis.}
\begin{tabular}{c c}\hline
Interval Name &Frequency range\\
 & (mHz)\\ \hline\hline
1st $p$-band&2.5 $<\nu<$ 3.5 \\ \hline
2nd $p$-band&3.5 $<\nu<$ 4.5 \\ \hline
3rd $p$-band&4.5 $<\nu<$ 5.5 \\ \hline
1st HIPs band&5.8 $<\nu<$ 6.3 \\ \hline
2nd HIPs band&6.3 $<\nu<$ 6.8 \\ \hline
\end{tabular}
\end{center}
\label{tab:tab1}
\end{table}
When one studies the solar oscillations spectrum with helioseismic instruments  the convective background (granulation, mesogranulation, supergranulation) and the effects of active regions crossing through the visible solar disk determine the noise background and hence limit the sensitivity of our measurements \citep[e.g.][]{Lef08}. There is also the impact of time gaps in the dataset. We next discuss our strategy for limiting their effect. Our main concern was to accurately determine the yearly averaged velocity amplitude in each of the five frequency bands. Instead of taking the data as a single, year-long dataset with gaps we chose to work with a large number of subsets, each of which has very high data fill of over 70 per cent. This reduces artefacts and improves the accuracy of the mean velocity amplitude. The choice of the subset length was very carefully considered. At low frequency we found that a subset of the length of a half a day was the best compromise between improving accuracy and maintaining sufficient resolution in the acoustic power spectrum \citep{Sim04}.
Hence, to track the solar cycle changes in the velocity or intensity amplitudes within the $p$-band, we split the timeseries for each year  into subsets with a length of half a day. We performed a Fast Fourier Transform (FFT) for every half day subset and then we integrated the velocity amplitudes in each of the five frequency intervals. Finally we averaged the 730 subsets to get the yearly mean velocity amplitude \citep{Sim09a}. The errors associated to each yearly mean velocity amplitudes were determined by taking the yearly standard deviation.
\subsection{Mean Fourier transform at high frequency}
\begin{figure*}
\centering
\mbox{\subfigure{\includegraphics[width=3in]{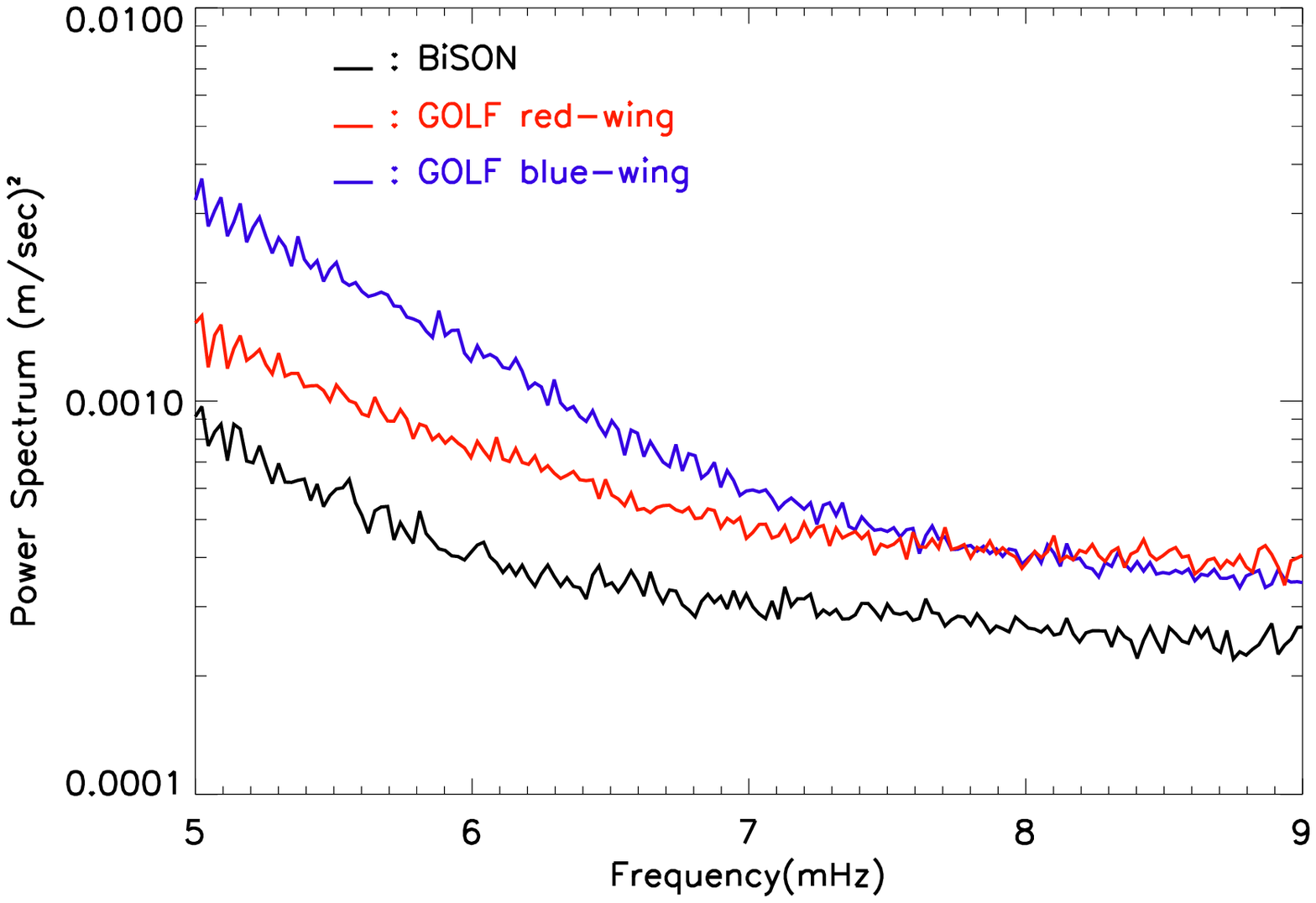}}\quad
\subfigure{\includegraphics[width=3in]{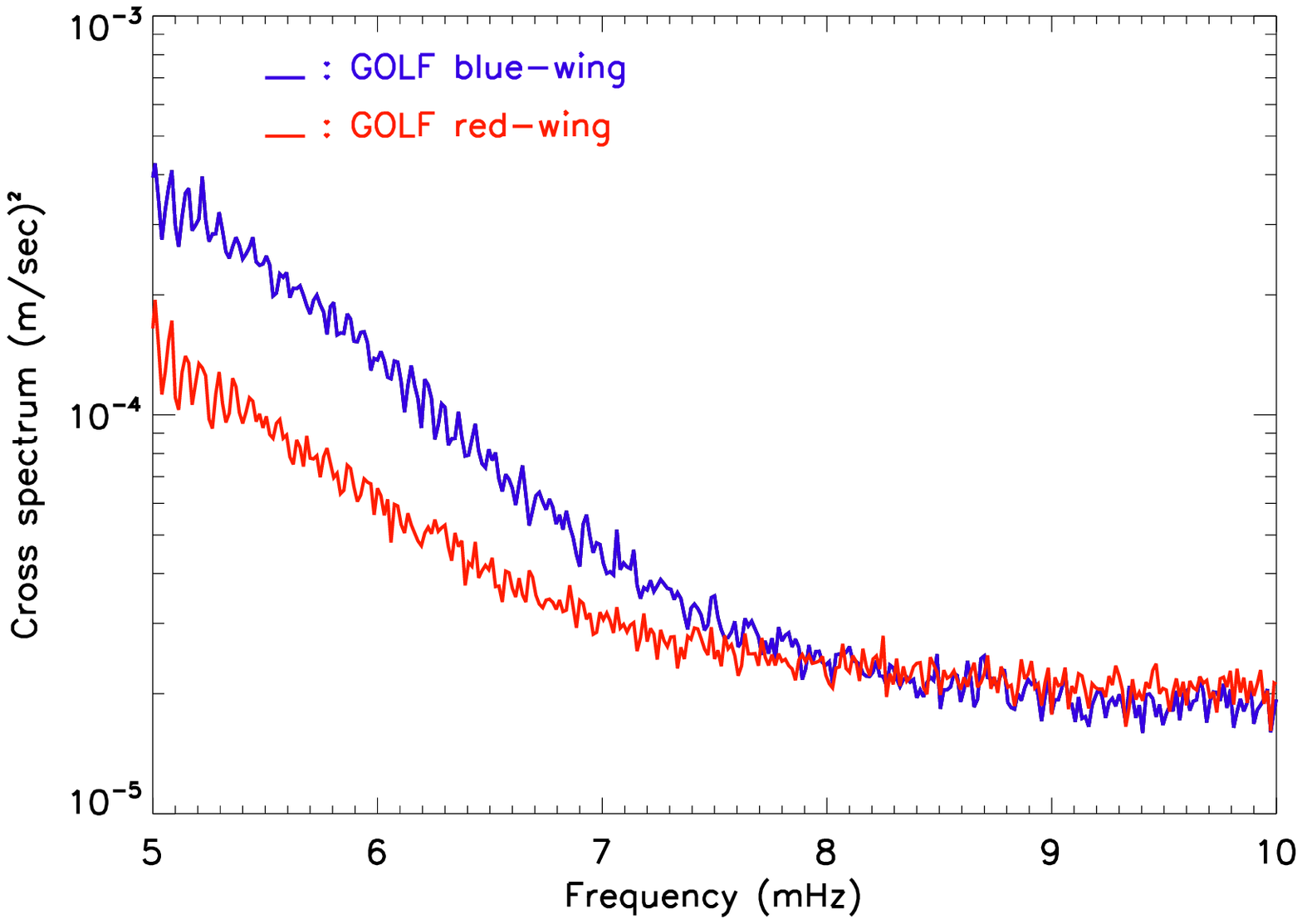} }}
\caption{Left-hand panel: BiSON and GOLF power spectrum in the HIPs band. Right-hand panel: GOLF cross-spectrum for the blue and red-wing observations in the HIPs band.} \label{fig:cross_golf}
\end{figure*}

The strongest signals detected in integrated sunlight observations are from standing waves whose amplitude results from the coherent addition of waves in the solar cavity. Traveling waves, however, seldom produce a coherent signal in time, and therefore we would not expect to find any such signal in the power spectrum of integrated sunlight. Despite this, a pattern of equally spaced peaks was found in GOLF, BiSON and VIRGO above the acoustic cut-off \citep{Gar98b,Cha03,Jima05}. They were named {\bf H}igh-frequency {\bf I}nterference {\bf P}eak{\bf s} (HIPs), and it is suggested that this high-frequency spectrum is the result of geometric interference between traveling waves \citep{kum90,kum91,Gar98b}. These waves have a much lower q-factor than is observed for the peaks around 3~mHz.
It has been shown that the best subset length to observe the HIPs is four days. Hence
we analyzed the data by splitting a yearly time series into subsets of four days' length. We performed the FFT for each subset and  then averaged them. Figure~\ref{fig:cross_golf} shows the high frequency part of the power spectrum from the GOLF blue and red-wing configuration and BiSON, where the color with which the GOLF data are shown is indicative of the observing wing. The three different signals have different amplitudes due to the different observational heights \citep{Sim08}. In this spectrum we can distinguish three different regions:

- between 5.0~mHz $<\nu<$ 5.5~mHz unresolved high-$n$ $p$-mode envelopes (odd and even degree) separated by 60 to 80~$\mu$Hz depending on the frequency interval;

- between 5.3~mHz $<\nu<$ 5.8~mHz mainly no signal for GOLF blue wing configuration, few peaks for BiSON and GOLF red-wing configuration;

- above 5.8~mHz in the GOLF blue-wing configuration up to $\approx$ 7~mHz a HIPs pattern that consists of equally spaced peaks of almost 70~$\mu$Hz. In the GOLF red-wing configuration and BiSON we can spot few peaks.
 
 It has been shown that at higher frequency the noise in GOLF observations has an instrumental origin \citep{Gar05,Tur04} and is dominated by photon shot noise. The red-wing configuration has a low  photon count compared with that in the blue wing and we expect the noise levels to be commensurately higher. This may be an explanations for the different behavior between the blue and red-wing configurations.

\subsection{Averaged cross-spectrum at high frequency}
We analyzed the GOLF data with the cross-spectrum technique to extract the common signal with a better signal-to-noise ratio. The averaged cross spectrum (CS) is defined as the product of  the Fourier transform of one channel multiplied by the complex  conjugate of the Fourier transform of the second one and then averaged over all the sub-series. It enhances any coherent signals between the two data sets \citep{Gar98b,Gar99}. We can apply the cross spectrum technique to the GOLF data, because the instrument has two photo-multipliers for each channel in the red and blue-wing configuration. Therefore we can determine the cross-spectrum by using two time series from the two photo-multipliers in the blue and red-wing configuration. Figure~\ref{fig:cross_golf} shows the cross-spectrum for the blue and red-wing configuration: 
the noise levels have been reduced  by a factor of better than 10  and the signal-to-noise ratio has improved by a factor of 2. The CS will improve the determination of the HIPs' amplitude variation with solar cycle. The lack of two simultaneous time-series prevents us from using BiSON data in the HIPs band. Therefore we will apply the cross-spectral analysis only to the GOLF observations in the HIPs band \citep{Sim09b}.

\section{Frequency dependence in velocity and intensity amplitudes with solar cycle}
\subsection{Solar cycle changes in $p$-mode velocity and intensity amplitudes}
We analyzed BiSON, VIRGO and GOLF data starting from 1996 June 25 until 2006 June 25 for BiSON and VIRGO, while for GOLF we stopped in 2002. As indicated earlier, we are still working on the best calibration of the third GOLF velocity segment (blue-wing configuration) and do not use the data because the absolute value of the power could be changed by the calibration. For BiSON and VIRGO we took the year 1996 as reference level of activity and will show the results over the full solar cycle for the three different frequency bands listed in Table 1. For GOLF we made different choices. Because it is observing at two different heights in the atmosphere, for the blue period we determined the amplitude variation with respect to 1996 and for the red-wing configuration with respect to 1998. The sensibility to the visible solar disk is also different between both wings as it was theoretically studied \citep{Gar98a, Hen99}.

\begin{figure*}
\centering
\mbox{\subfigure{\includegraphics[width=3in]{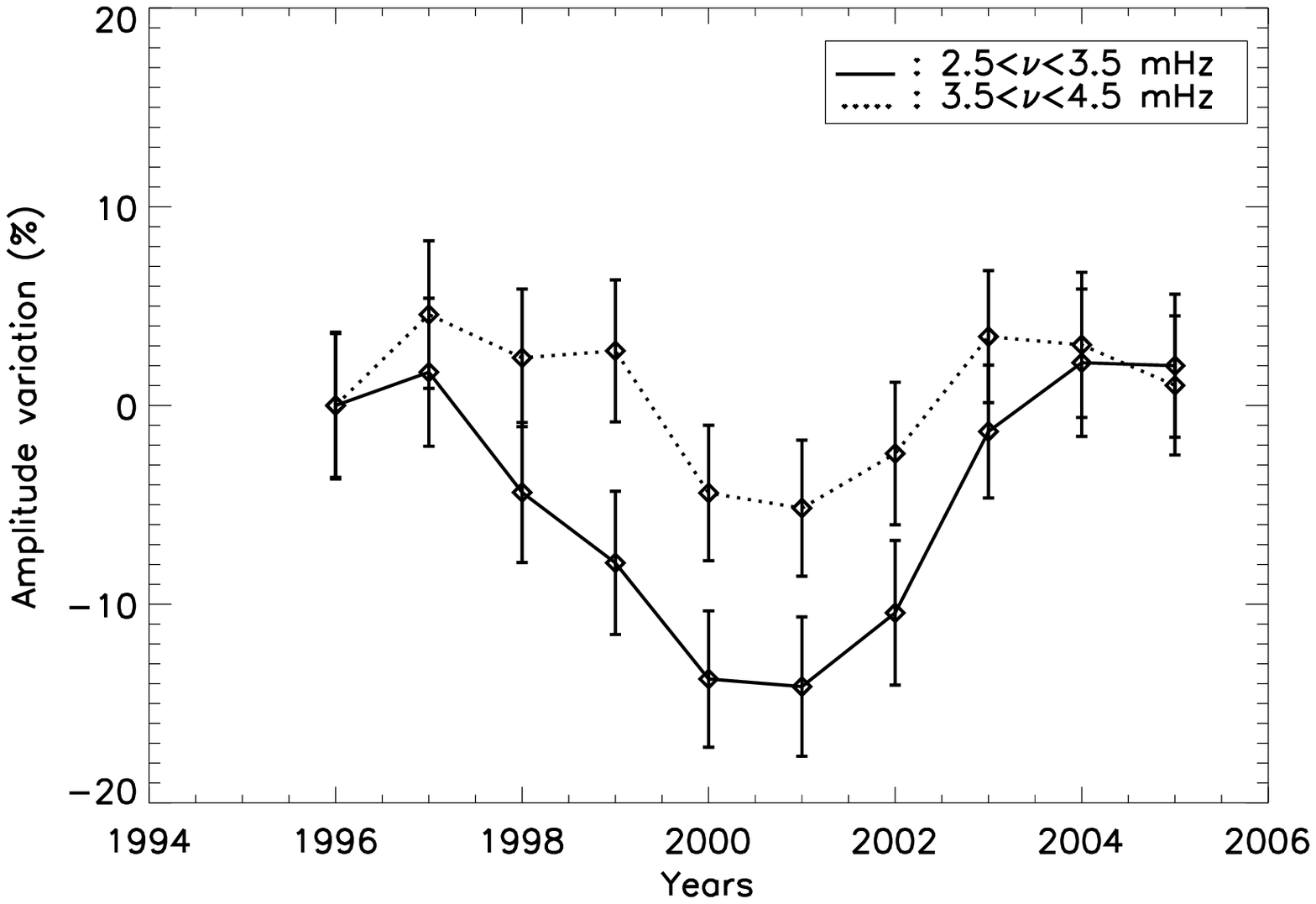}}\quad
\subfigure{\includegraphics[width=3in]{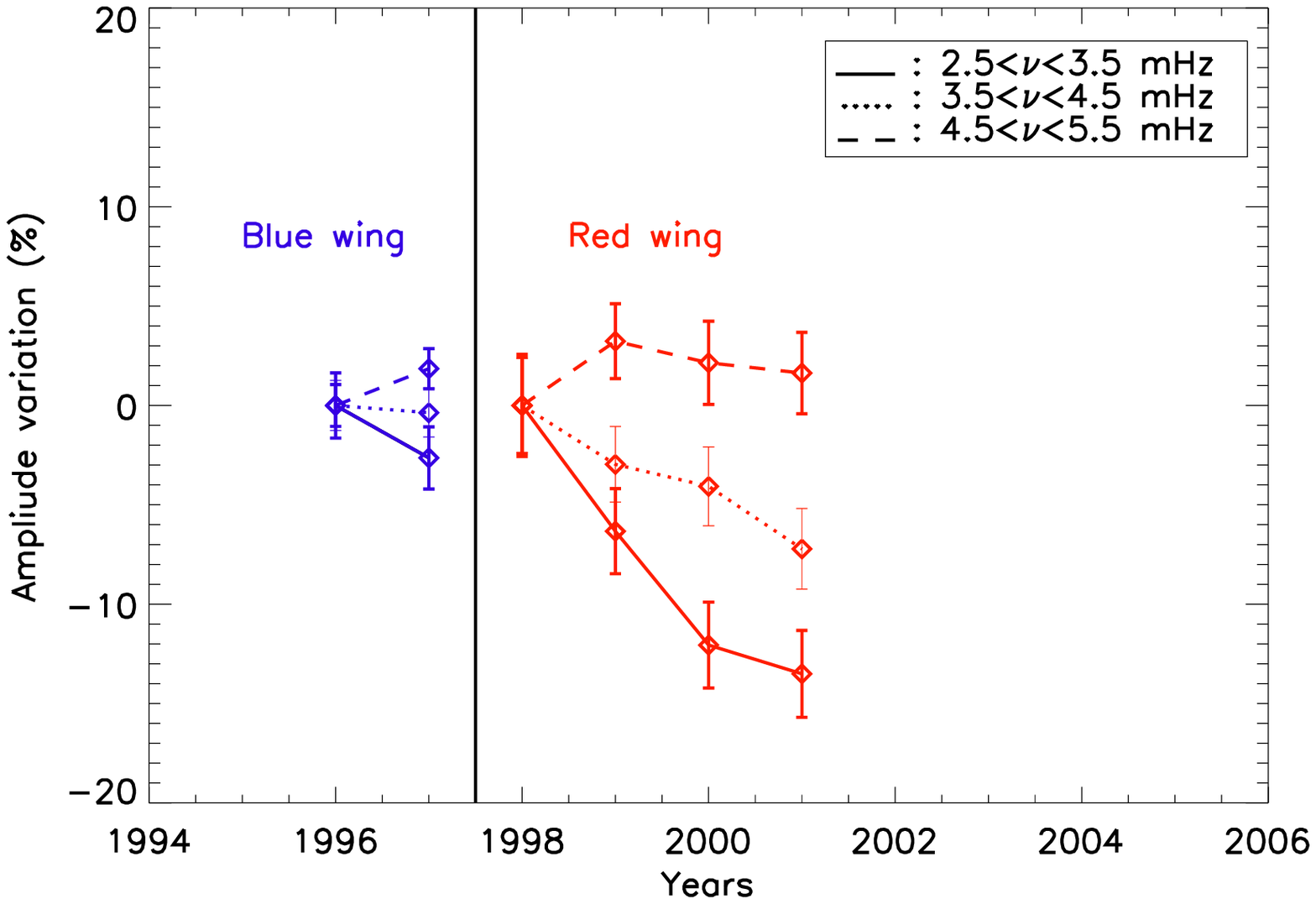} }}
\caption{Left-hand panel: solar cycle changes in the $p$-bands from the BiSON observations. Right-hand panel: solar cycle changes in the $p$-bands from the GOLF blue and red-wing observations.} \label{fig:cycle_lf_vel}
\end{figure*}

Figure~\ref{fig:cycle_lf_vel} shows velocity amplitude changes with solar cycle from the BiSON (left panel) and GOLF (right panel) observations. It shows several interesting features, and Table 2 quantifies the maximum value of the solar cycle changes for the different frequency bands:
\begin{table*}
\begin{center}
\caption{Maximum amplitude of the $p$-mode power variation as seen from different observational programs. Positive values indicate enhancement while negative values indicate suppression to the reference level.}
\begin{tabular}{c c c c c c}\hline
  & \multicolumn{5}{c}{Maximal Power Variation ($\%$)}\\
Interval&BiSON&GOLF&\multicolumn{3}{c}{VIRGO}\\
Name&&RED&RED&GREEN&BLUE\\ \hline
1st $p$-band&-14$\pm$ 3&-13$\pm$ 2&-10$\pm$ 2&-9$\pm$ 2&-8$\pm$ 2\\ \hline
2nd $p$-band&-4$\pm$3&-7$\pm$2&-1$\pm$2&-1$\pm$2& -1$\pm$2\\ \hline
3rd $p$-band&//&2$\pm$2&1$\pm$2&0$\pm$2& 3$\pm$2\\ \hline
\end{tabular}
\end{center}
\label{tab:tab2}
\end{table*}

- BiSON in the first frequency band shows a net decrease of about 14$\pm 3\%$. Within the errors, GOLF observations also show a suppression of about 13$\pm 2\%$ that agrees well with BiSON. Furthermore these findings agree well with previous results \citep{Cha00,Kom00,Jim04}

- in the second frequency band the $p$-mode amplitude suppression is sharply reduced compared to the 1st frequency band for both observational programs;

- in the third frequency band we find almost null change in the sign of the variation from the GOLF observations. For BiSON we cannot provide any estimate of the solar cycle changes in this band due to an instrumental artefact that dominates any solar cycle changes induced in $p$-mode velocity amplitudes.

\begin{figure*}
\centering
\mbox{\subfigure{\includegraphics[width=2.3in]{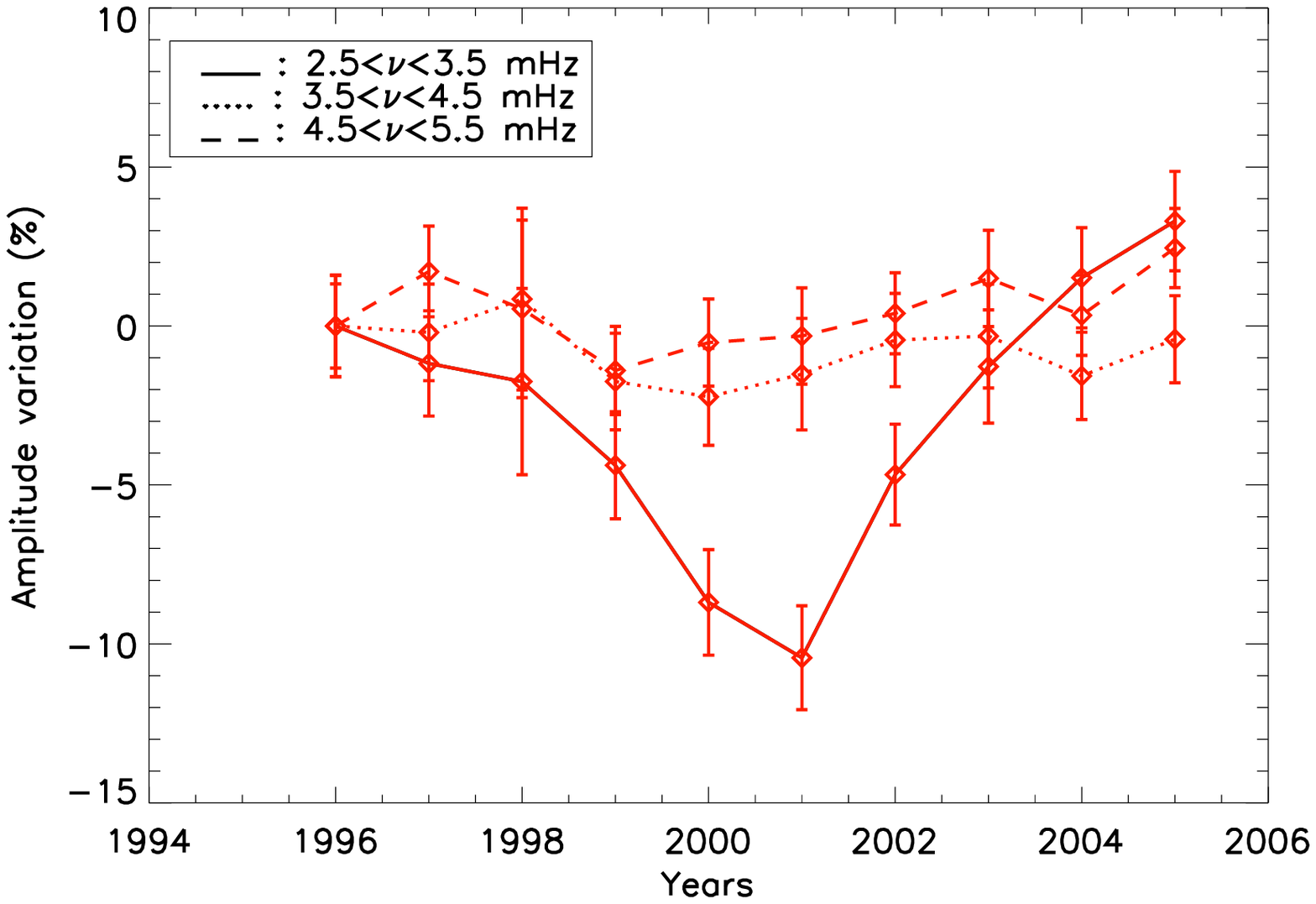}}\quad
\subfigure{\includegraphics[width=2.3in]{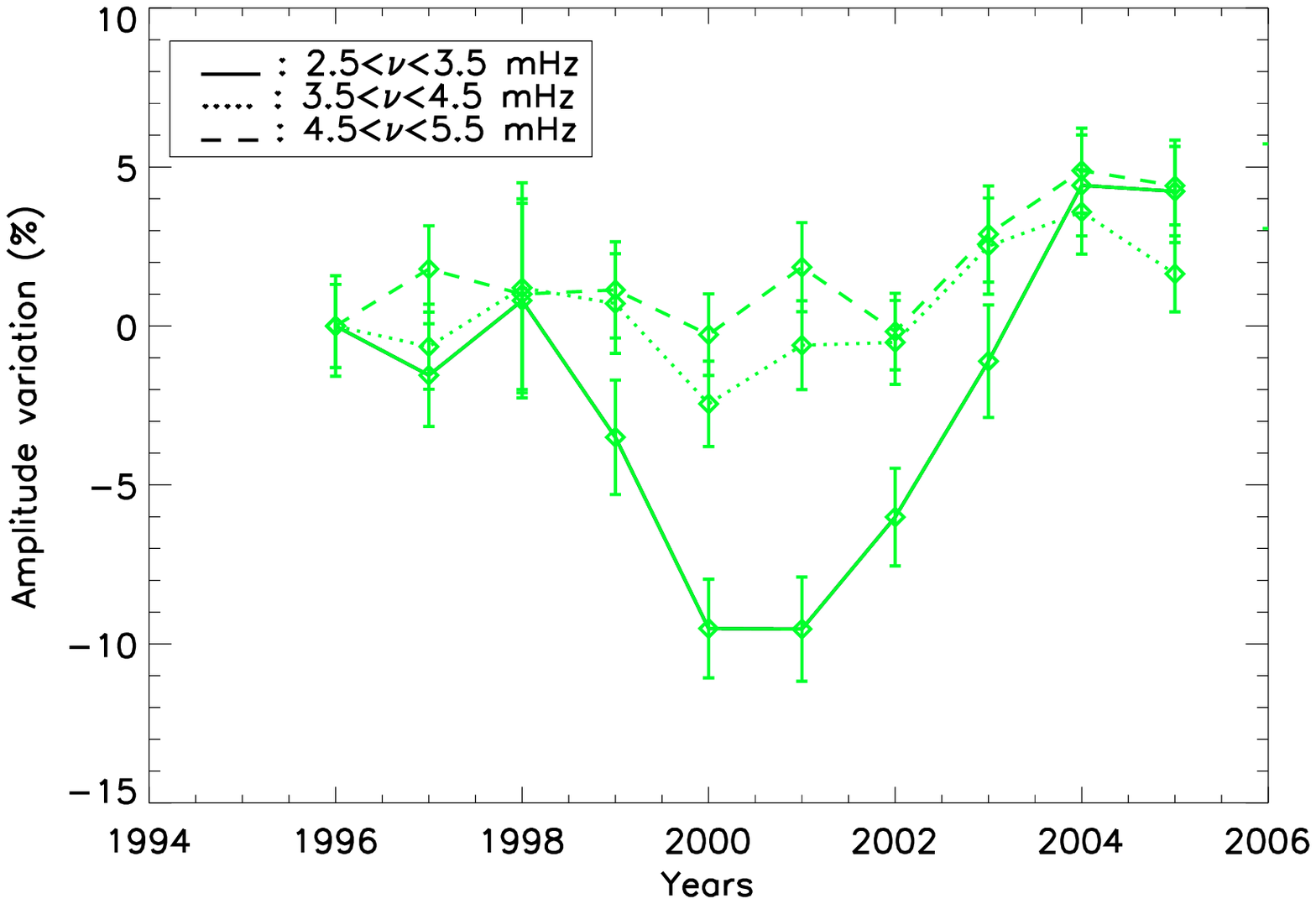}}\quad
\subfigure{\includegraphics[width=2.3in]{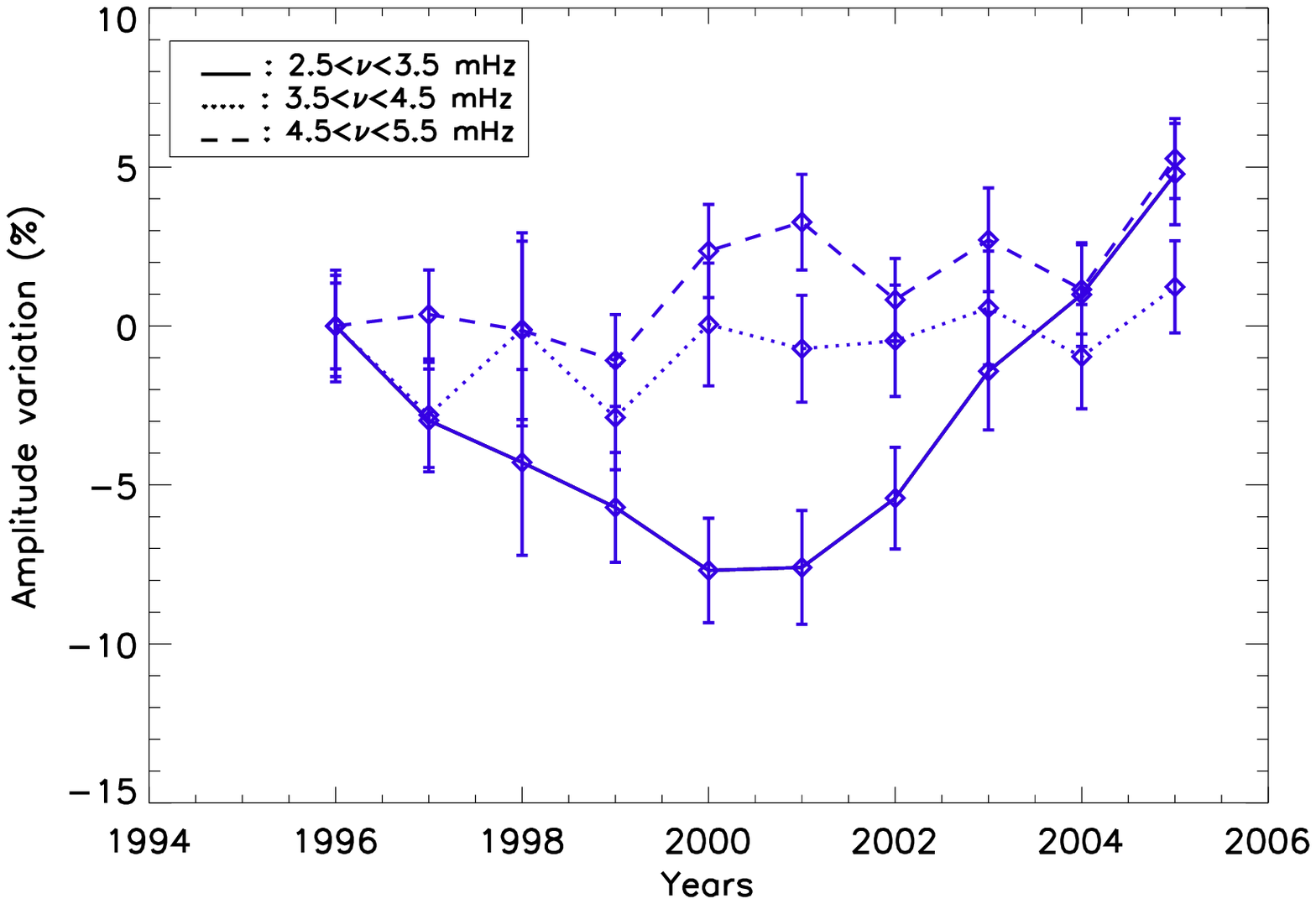}}}
\caption{Solar cycle changes in the $p$-bands from the VIRGO red (left), green (center) and blue (right) observations.} \label{fig:cycle_lf_int}
\end{figure*}

We also used VIRGO observations to investigate $p$-mode amplitude variation in intensity observations. Figure~\ref{fig:cycle_lf_int} shows $p$-mode amplitude variation from intensity observations:

- in the first frequency band we found a $p$-amplitude suppression of 9$\pm 2\%$ from all the three channels. Their values indicates a slight weaker suppression in this frequency band compared to the findings in the velocity observations;

- in the second frequency band the variation is slightly negative;

- in the third frequency band we find a null change.

The findings obtained from the velocity observations highlight a frequency dependence of the size of the suppression over the ascending phase of solar cycle 23. Indeed the findings shows a maximal suppression in the first $p$-band, which becomes less pronounced in the second and vanishes in the third $p$-band. In intensity observations, we found the maximal suppression in the first $p$-band, but in the second and third the size of the suppression is almost null in both $p$-bands (Table 2). Nevertheless the common features between velocity and intensity observations is that the maximal suppression is observed in the first $p$-band and an almost null suppression in the third $p$-band.

\subsection{$p$-mode correlation with actvity proxy}
Velocity observations seem to suggest a solar cycle dependence of the size of the suppression for two out of three with the progression through solar cycle. Therefore we decided to analyze the yearly amplitude changes as a function of two well-known proxies of global surface activity: the International Sunspot Number (ISN) and MgII H and K core-to-wing ratio. We used these two activity proxies, because they are sensitive to different aspects of the  magnetic flux. The ISN is sensitive only to the strong polarized magnetic flux component, while the MgII core-to-wing-ratio is sensitive to the entire solar disk and therefore also includes the contribution from bright faculae and network \citep{Ver01}. Furthermore was shown that the MgII activity proxy provides the most consistent description of $p$-mode frequency shifts over three solar cycles \citep{Cha07}. Averages of each proxy were computed over the same periods over which the data were analyzed to derive each of the $p$-mode yearly mean velocity amplitudes. Due to the agreement between BiSON and GOLF in the first two $p$-bands and because ppression in the VIRGO observations is almost null already in the second $p$-band (we will explain the reason in sec. 5.1), we decided to calculate the correlation between the GOLF observations and the two activity proxies.
\begin{figure*}
\centering
\mbox{\subfigure{\includegraphics[width=3in]{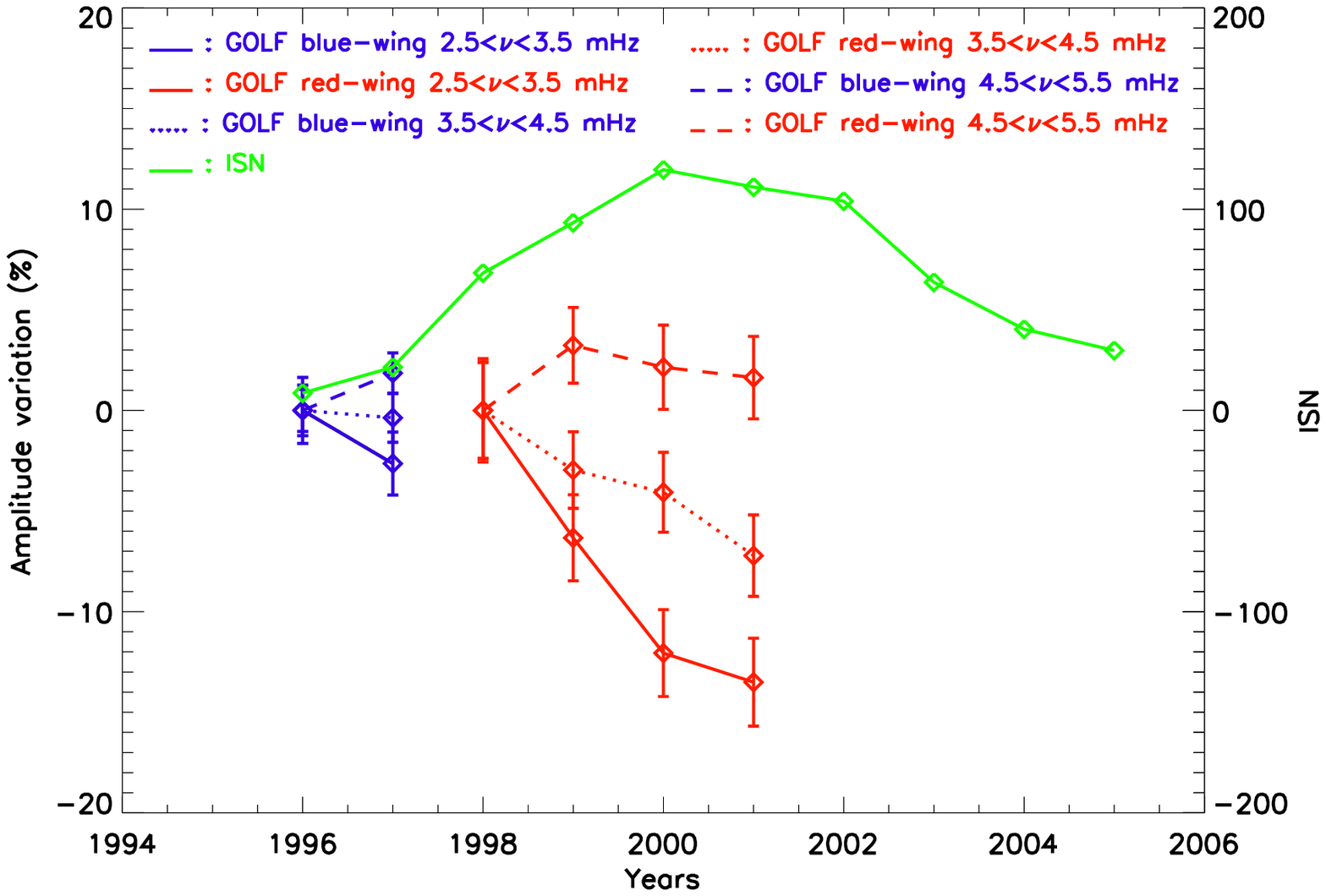}}\quad
\subfigure{\includegraphics[width=3in]{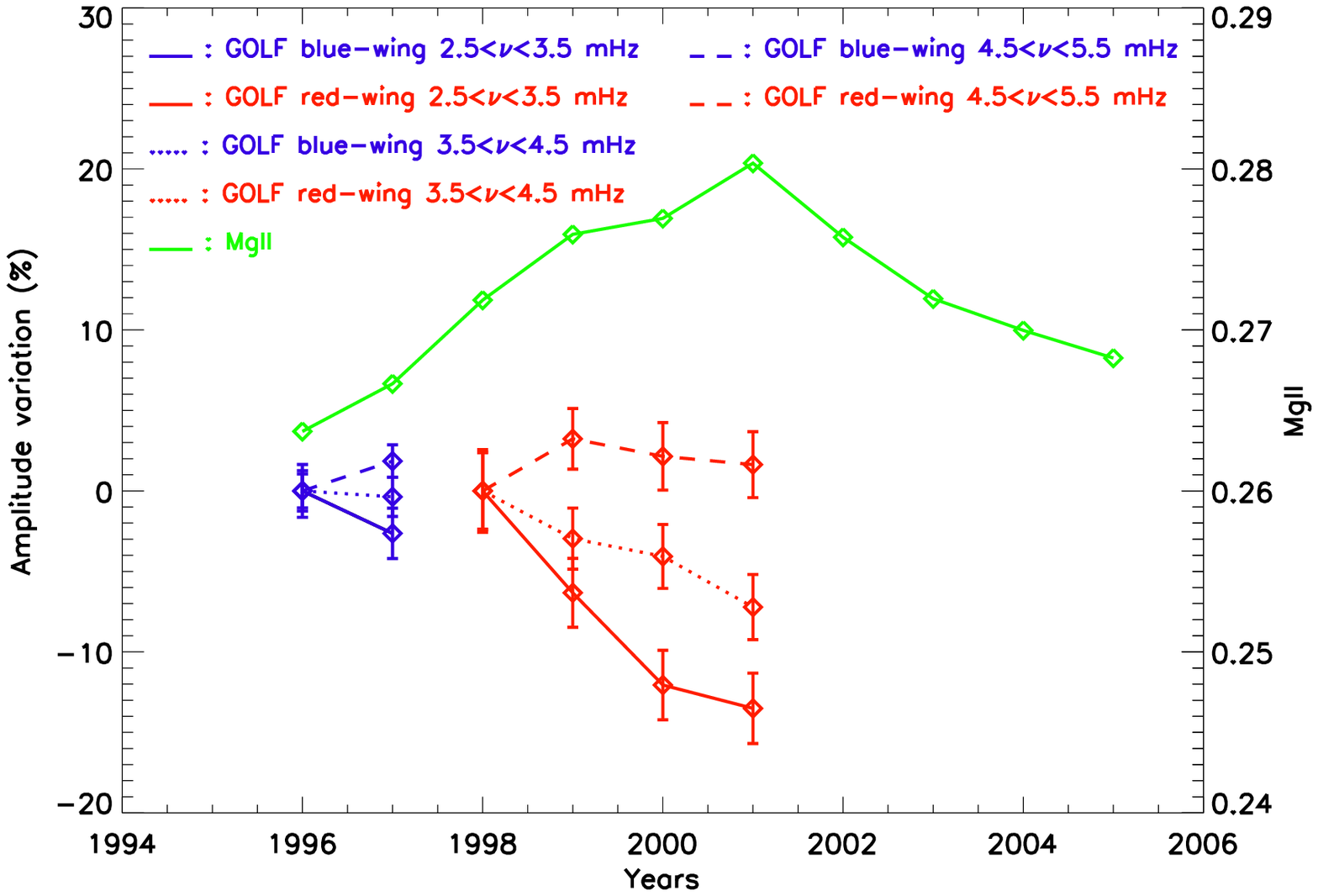} }}
\caption{ Solar cycle changes in the $p$-bands from the GOLF blue and red-wing observations compared to the ISN (left-hand panel) and MgII (right-hand panel). In green the ISN (left panel) and MgII (right hand panel) activity proxies.} \label{fig:golf_act}
\end{figure*}
Figure~\ref{fig:golf_act} shows $p$-mode velocity amplitudes variation as seen by GOLF red-wing configuration over the period 1996-2001 compared to the changes observed in ISN and MgII.
It clearly shows that the size of the $p$-mode velocity amplitude suppression increases with increasing level of magnetic activity for both activity proxies in two out of three $p$-bands,. We then calculated the correlation coefficient with the Spearman rank formula, because it also provides the corresponding probability of a null correlation.
\begin{table}
\begin{center}
\caption{Spearmann rank coefficient between the yearly amplitude variation in the three $p$-bands and the two activity proxies ISN and MgII.}
\begin{tabular}{c c c c c}\hline
Interval&\multicolumn{2}{c}{ISN}&\multicolumn{2}{c}{MgII}\\
Name& Corr & Rk&Corr & Rk \\ \hline
1st $p$-band& -80&20&-80&20\\ \hline
2nd $p$-band& -80&20&-80&20\\ \hline
3rd $p$-band& 40& 60&40&60\\ \hline
\end{tabular}
\end{center}
\label{tab:golf_lf_sn}
\end{table}
Table 3 indicates a correlation of 80$\%$ in two out of the three $p$-bands for ISN and MgII activity proxies. These findings agree with previous results \citep{Jim04,Cha07}.

\subsection{Solar cycle changes in the HIPs' velocity and intensity amplitudes}
We used 11 years of the VIRGO data starting from 1996 June 25 until 2006 June 25, while for the GOLF observations we stopped in 2002. We split the whole time series in subset length of four days to observe the HIPs. We performed the averaged cross-spectrum analysis by taking two time series in the blue and red-wing configuration of the GOLF observations. Then we performed the CS in each of the individual subsets and averaged the CS over one year. We considered the HIPs frequency band between 5.8~mHz $<\nu<$ 6.8~mHz in two parts each of the width 0.5~mHz and present the yearly averages of the signal in these bands in Table 4.

\begin{figure*}
\centering
\mbox{\subfigure{\includegraphics[width=3in]{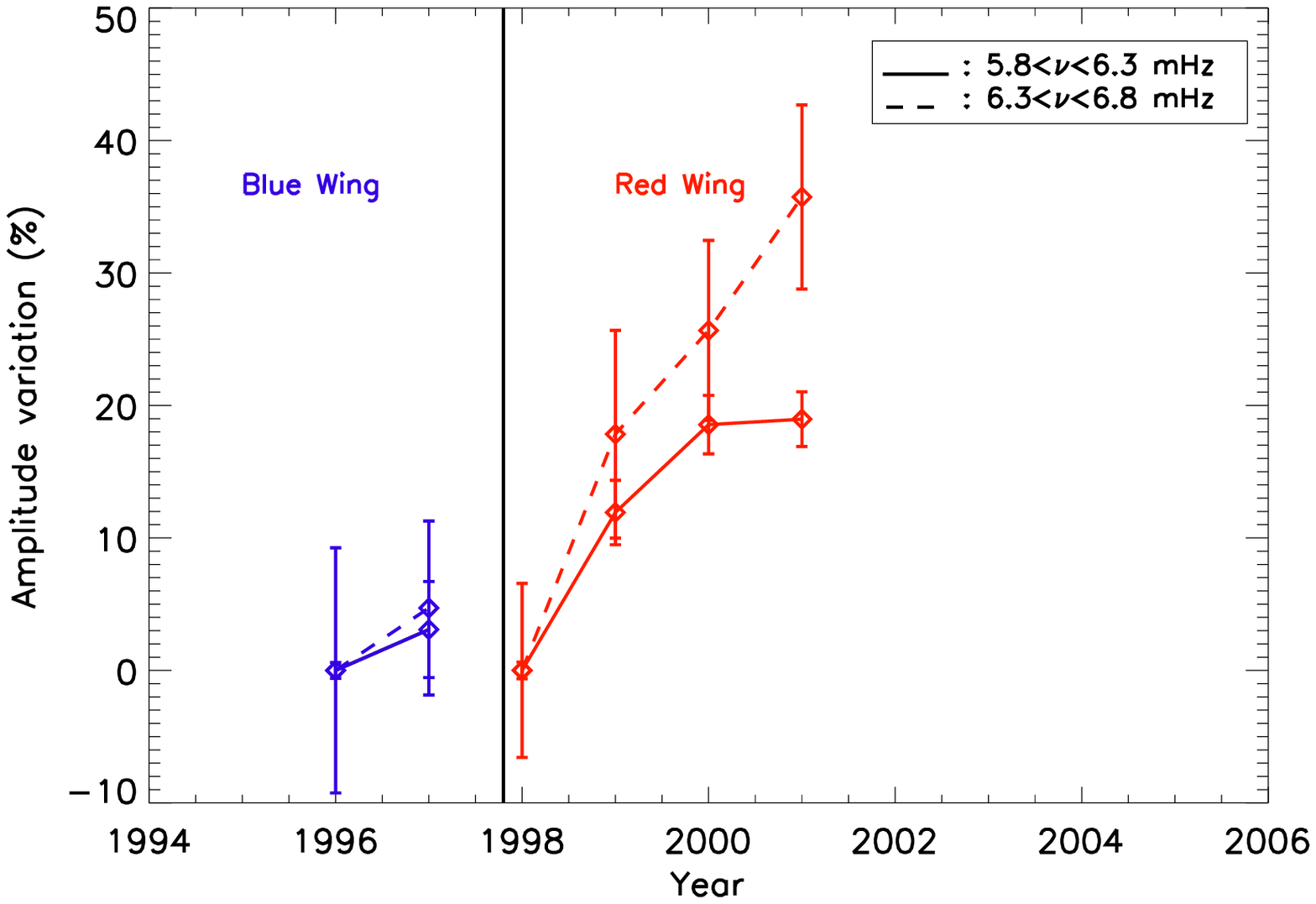}}\quad
\subfigure{\includegraphics[width=3in]{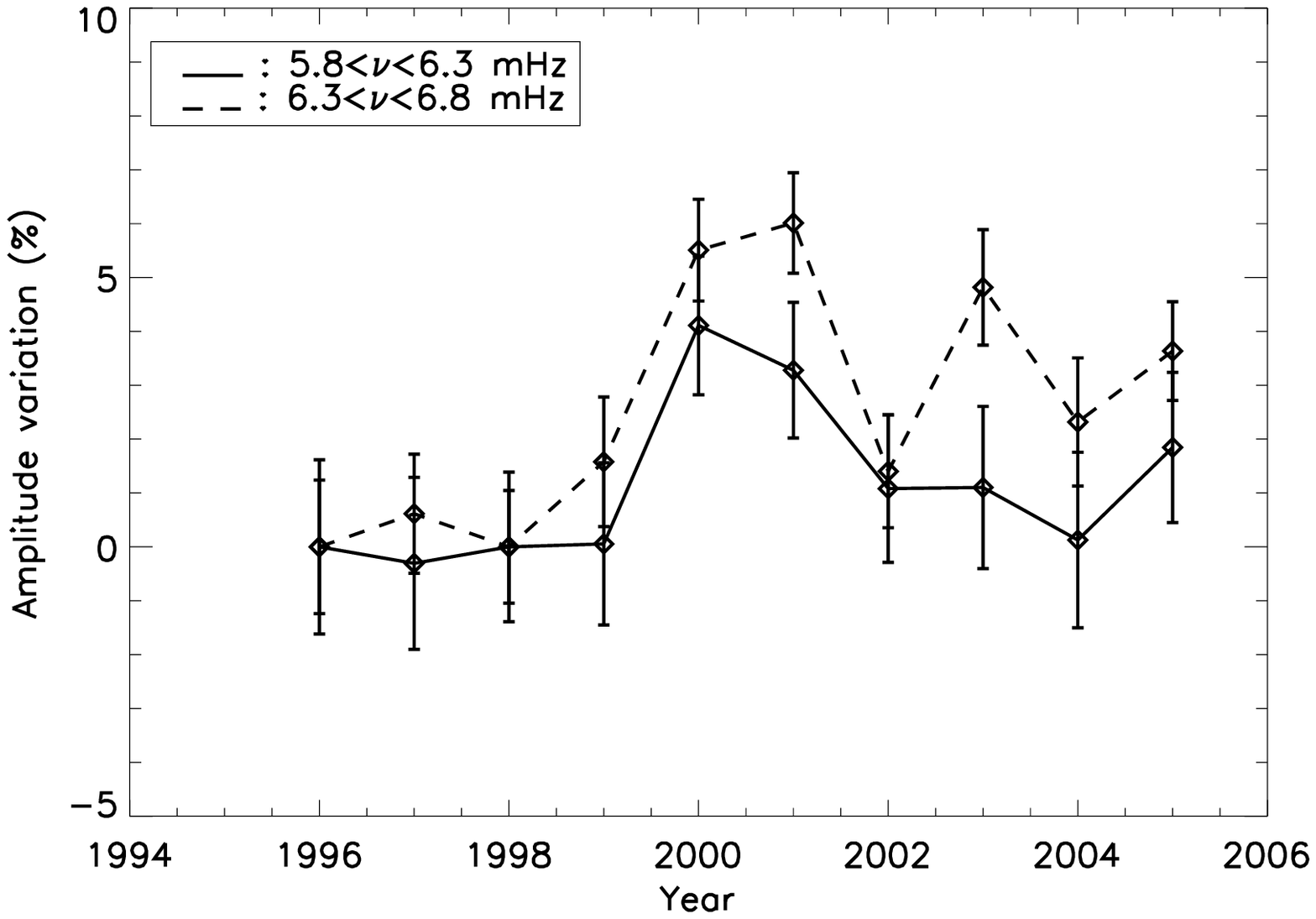} }}
\caption{Left-hand panel: solar cycle changes in the HIPs bands from the GOLF blue and red-wing observations. Right-hand panel: solar cycle changes in the HIPs bands from the VIRGO observations.} \label{fig:cycle_hf}
\end{figure*}

Figure~\ref{fig:cycle_hf} shows the HIPs velocity amplitude variation over solar cycle 23 for the GOLF and VIRGO observations. We also determined the amplitude vaiation for the blue period in the same way as for the lower frequency band, because GOLF observes at two different heights in the atmosphere, that is, we determined the blue period with respect to the data from 1996 and those of the red-wing configuration with respect to those of 1998.
We found an increase of acoustic emissivity of about ~18$\pm$3 per cent over the period 1998-2001 for the first HIPs band, while $\approx$~36$\pm$7 in the second HIPs band. This finding seems to point to an increase of acoustic power with increasing frequency.
Based on these results, we decided to look for solar cycle changes in the HIPs band by using the continuum intensity data provided by the VIRGO Sun photometers. The signal-to-noise ratios are poor and hence we decided to apply the CS technique to the various different color channels of the VIRGO observations, in order to enhance any coherent signal.  We improved the signal-to-noise ratio by a factor of 2.

Figure~\ref{fig:cycle_hf} (right panel) shows the HIPs intensity amplitude variation over solar cycle 23 between the blue and red channels.
As we can see we find a slight indication of an increasing acoustic emissivity over the solar cycle within the errors, but the size of the enhancement is pretty small. It is important to underline that the CS analysis performed between blue and green and green and red shows up with a pretty similar behavior as the blue and red channel.
\begin{table}
\begin{center}
\caption{Maximum amplitude of the HIPs' power variation as seen from the GOLF and VIRGO observations. Positive values indicate enhancement to the reference level.}
\begin{tabular}{c c c}\hline
  & \multicolumn{2}{c}{Maximal Power Variation ($\%$)}\\
Interval&GOLF&{VIRGO}\\
Name&RED&RED-GREEN\\ \hline
1st HIPs band&18$\pm$3&3$\pm$2 \\ \hline
2nd HIPs band&36$\pm$7&5$\pm$2\\ \hline
\end{tabular}
\end{center}
\label{tab:tab_golf_hips}
\end{table}
\subsection{HIPs correlation with actvity proxy}
 The findings from the GOLF red-wing configuration seems to point to an increasing acoustic power with increasing magnetic activity cycle, and the size of the effect increases with frequency. Although, we have only four points, we decided to analyze the yearly mean amplitude changes in the two HIPs bands as a function of our chosen proxies of global surface activity: the ISN and and MgII H and K core-to-wing ratio.
\begin{figure*}
\centering
\mbox{\subfigure{\includegraphics[width=3in]{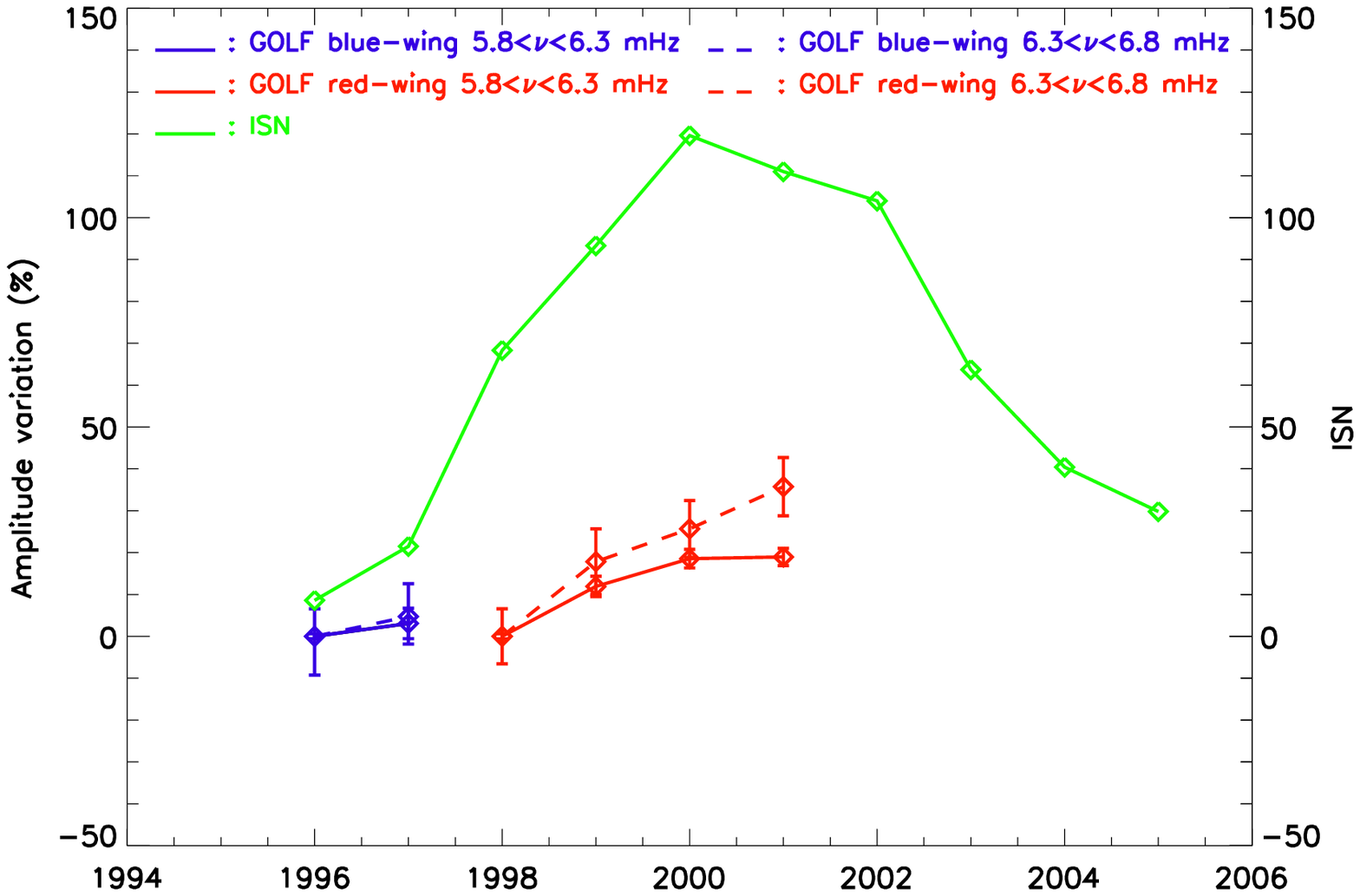}}\quad
\subfigure{\includegraphics[width=3in]{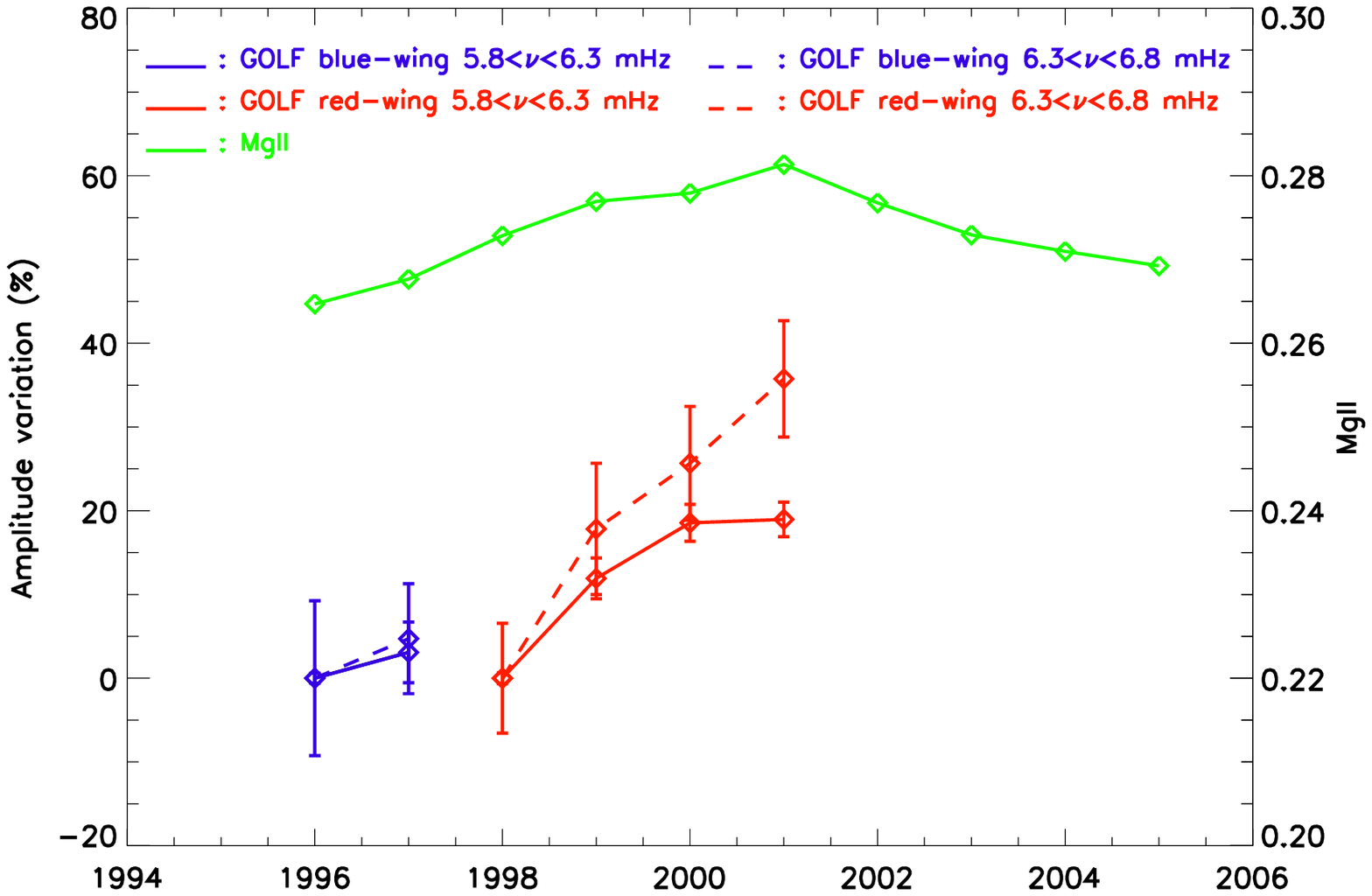} }}
\caption{Solar cycle changes in the HIPs bands from the GOLF blue and red-wing observations compared to the ISN (left-hand panel) and MgII (right-hand panel). In green the ISN (left panel) and MgII (right hand panel) activity proxies.}
 \label{fig:golf_mgII}
\end{figure*}
Figure~\ref{fig:golf_mgII} shows the amplitude variation in the two HIPs bands compared to the ISN variation over solar cycle 23 in the left panel, while the right panel shows MgII. In both cases it is abundantly clear that the acoustic power increases with increasing solar magnetic activity. We calculated the Spearman rank coefficient as in Table 5. The results point to a high correlation with the ISN and a correlation of 1 with MgII. However, we should emphasize that  we used only four points out of 11 of the full solar cycle for the two determination.
\begin{table}
\begin{center}
\caption{Spearman rank coefficient between the yearly amplitude variation in the two HIPs bands and the two magnetic activity proxies ISN and MgII.}
\begin{tabular}{c c c c c}\hline
Interval&\multicolumn{2}{c}{ISN}&\multicolumn{2}{c}{MgII}\\
Name& Corr & Rk&Corr & Rk \\ \hline
1st HIPs band& 80&20&1&20\\ \hline
2nd HIPs band& 80&20&1&20\\ \hline
\end{tabular}
\end{center}
\label{tab:golf_hf}
\end{table}
\section{Discussion}
\subsection{Observational evidences of mode conversion}
\begin{figure*}
\centering
\mbox{\subfigure{\includegraphics[width=3in]{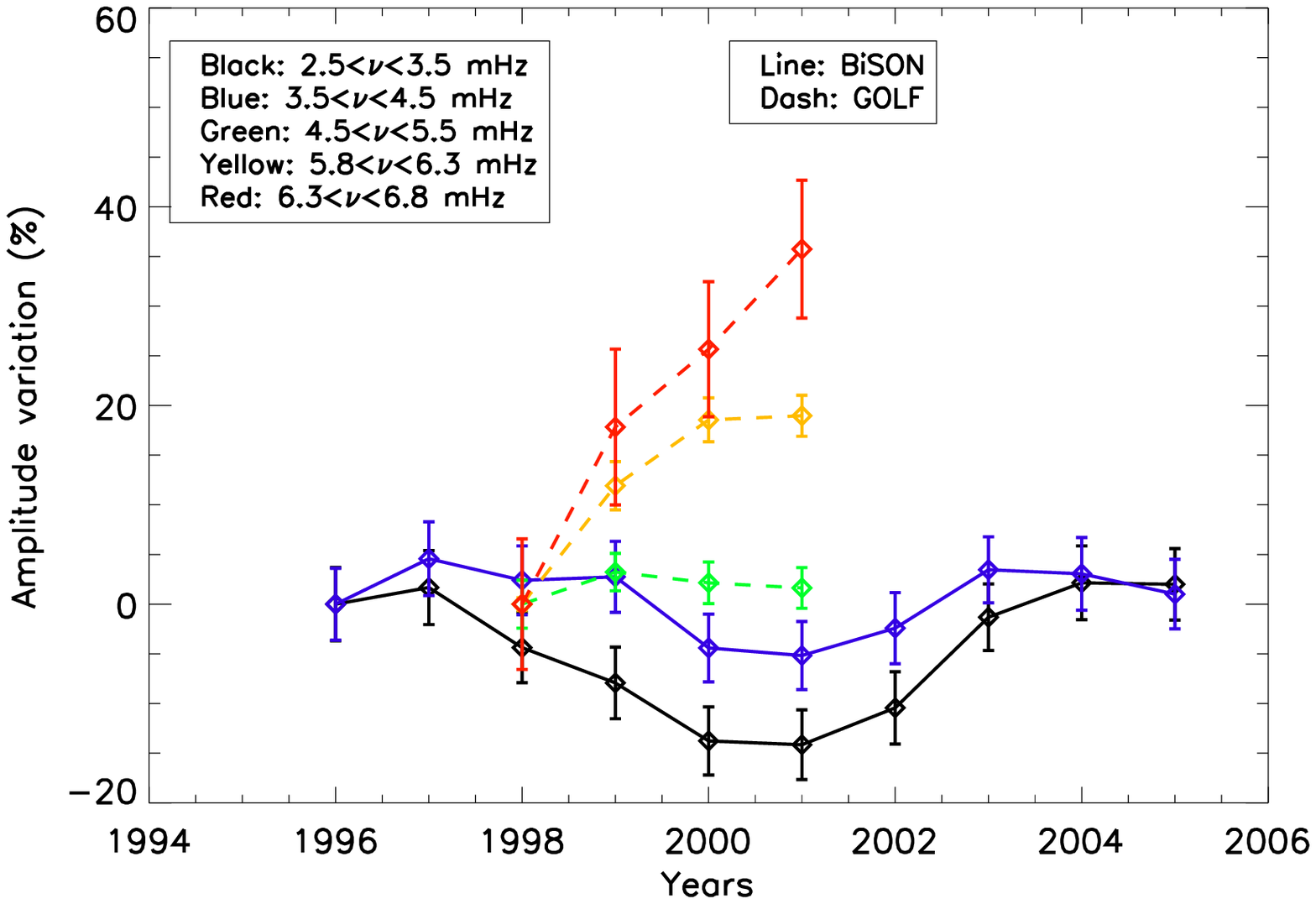}}\quad   
\subfigure{\includegraphics[width=3in]{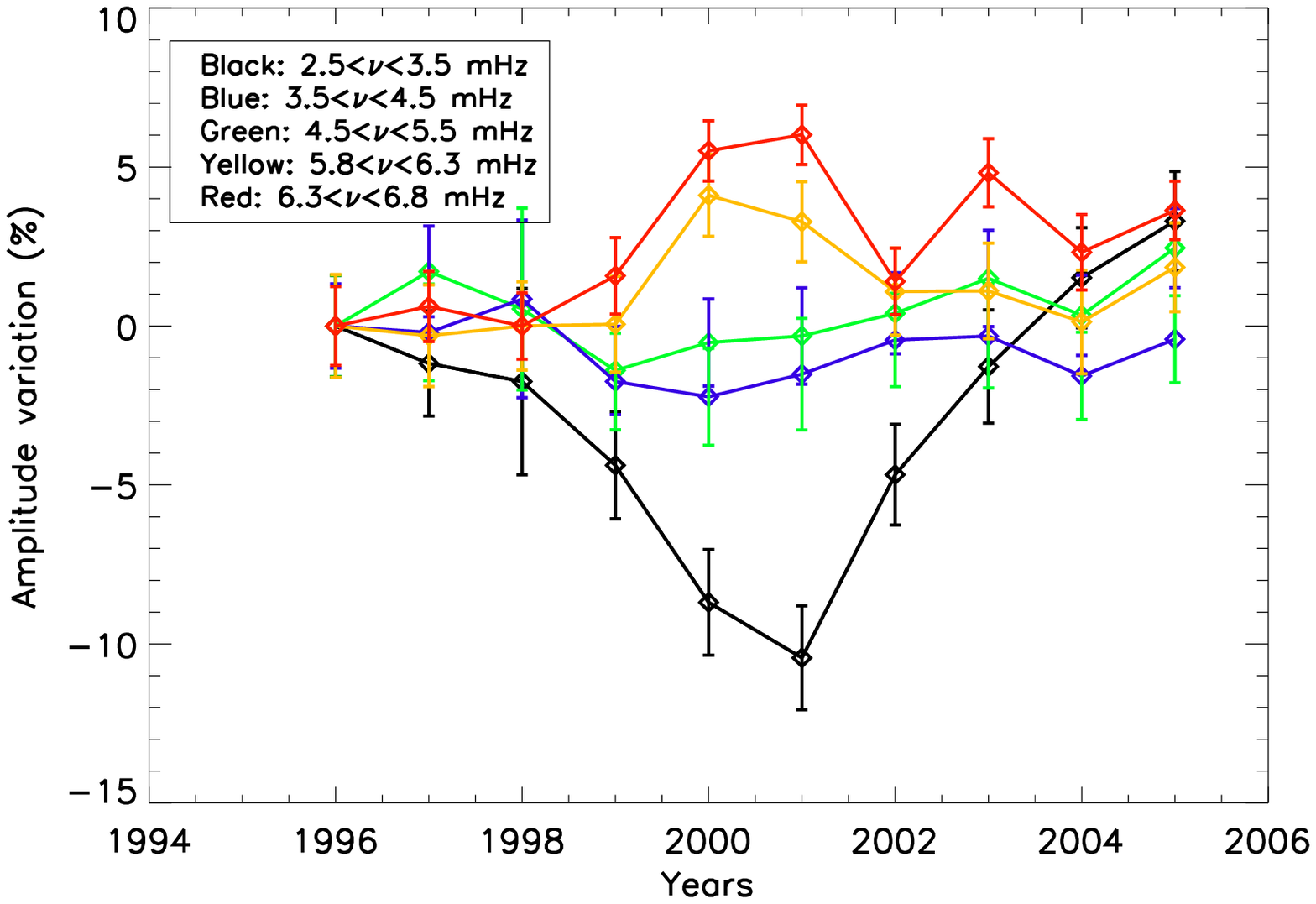} }}
\caption{Left-hand panel: solar cycle changes in the $p$ and HIPs bands from the BiSON and GOLF red-wing observations. Right-hand panel: solar cycle changes in the $p$ and HIPs bands from the VIRGO observations.} \label{fig:comp_vel}
\end{figure*}

We investigated solar cycle changes in acoustic power with frequency and with the progression through the solar cycle. In the $p$-bands we found an acoustic power suppression that strongly decreases with increasing frequency. Furthermore the GOLF and BiSON observations show a good agreement in the strength of the effect within the errors. In the HIPs bands we found instead an acoustic power enhancement with the solar cycle. Therefore we decided to assemble these findings in a {\it composite} plot as
Fig.\ref{fig:comp_vel} shows. It clearly visualizes the change in the sign of the variation with increasing frequency.
We built up the same composite plot by using intensity observations. It shows that for increasing frequency the acoustic enhancement observed in intensity data is pretty small.
 The same frequency dependence of $p$-mode velocity amplitudes was found in MDI observations, where $p$-mode amplitudes in magnetic regions and quiet Sun were compared \citep{Jai02}. We now consider an explanation for the variations.
 
\begin{figure}
\centering
\includegraphics[width=3in]{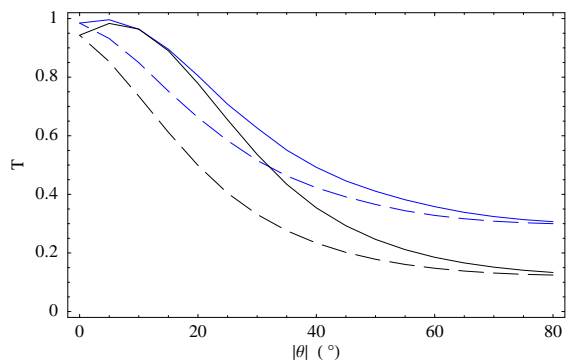}
\caption{The transmission coefficient, $T$,  \citep{sch06} against magnetic field
inclination, $\theta$, for a 6~mHz wave for a ray with a lower turning  depth
of 35~Mm. The blue curve is the 1~kG case and the black curve is the 2~kG case. The dashed
line is for the inclination away from the wave vector and the solid line is
for the inclination in the same direction as the wave vector. When $T$ is small
there is strong conversion of internal acoustic waves across the equipartition layer $z_{eq}$ to fast magnetoacoustic waves.}
\label{fig:tra}
\end{figure}

 Mode conversion has been proposed as a possible mechanism for describing the $p$-mode acoustic energy absorption in sunspots and active regions \citep{Cal03, Cro05}. A strong surface magnetic field allows the initial internal acoustic wave energy to be split into fast and slow magnetoacoustic branches near the $z_{eq}$ equipartition layer  where $a\approx c$. This splitting has to satisfy the following conservation wave energy: fast energy + slow energy = total energy. If the magnetic field is enough inclined to lower the acoustic cut-off frequency in low $\beta$ plasmas, the transmitted slow magnetoacoustic wave may propagate into the solar atmosphere (ramp effect). If the acoustic cut-off frequency has not been reduced, the slow magnetoacoustic wave is instead refracted backwards before reaching the observation height. The transmission coefficient, $T$ \citep{sch06}, is defined as the amount of energy transmitted from internal acoustic-to-slow or slow-to-internal acoustic waves during mode transmission. It has been shown that for the same magnetic field strength and inclination, $T$ is smaller for  higher frequencies \citep[fig. 15]{sch06}. This implies that (in the interior of the Sun) the transmission from internal acoustic waves to slow magnetoacoustic becomes less dominant with increasing frequency. Figure~\ref{fig:tra} shows the transmission coefficient, $T$, of a 6~mHz internal acoustic wave which transmits into the low-$\beta$ region as a slow magnetoacoustic wave in the generalized ray theory approximation by \citet{sch06}. For a 2~kG magnetic field strength and $60^\circ$ inclination the transmission coefficient is below $\approx 20\%$, while for a 1~kG magnetic field strength it is below $\approx 40\%$. Here the ray has a lower turning depth of 35~Mm, which corresponds to an  to harmonic degree $\ell \approx 300$ with an angle to vertical at the surface of less than $10^\circ$ \citep[fig.7]{Blo07}. This is not exactly comparable to the global modes case, where $\ell$ is very small, but it is the current depth limit for the code as it is set up for local helioseismology.  Extrapolation to more vertical waves  at the surface (and smaller $\ell$) brings the dashed and solid lines in Fig.~\ref{fig:tra} together.
Our observations found a maximum suppression of power between 2.5~mHz $<\nu<$ 3.5~mHz, possibly caused by slow magnetoacoustic waves traveling upwards. Therefore it may be the case according to the mode conversion theory that the field inclination is of the order of $60^\circ$ to allow the resonant modes to be transmitted predominantly as slow-magnetoacoustic waves traveling outwards from the photosphere. For higher frequencies between 3.5~mHz $<\nu<$ 4.5~mHz and for the same field inclination, the transmission coefficient is slightly smaller in agreement with the reduced acoustic power suppression found in the second $p$-band. In the third $p$-band the transmission coefficient gets even smaller which allows the resonant modes to be split almost equally between the slow and the fast component. This agrees with the almost null suppression found in the third $p$-band. It fails, however, to explain the enhancement of power in the HIPs bands.
Above the acoustic cut-off, at the $60^\circ$ field inclination, the internal acoustic waves above 6~mHz retain their fast nature. In this case, the two restoring forces acting to generate the fast magnetoacoustic mode (gas pressure and magnetic pressure fluctuations) are in phase and add constructively for the fast mode increasing the acoustic power. Therefore, we observe a change in the sign of the integrated acoustic power above the acoustic cut-off with a maximum occurring between 6.3~mHz$<\nu<$6.8~mHz.

 The weak point is the lack of enhancement in intensity observations. The possible explanation of the observed difference between intensity and velocity observations could be found in the position of the equipartition layer ($z_{eq}$). The HIPs are traveling  waves in nature, and therefore in order for mode transmission/conversion to occur the $z_{eq}$ layer has to lie above the excitation sources that are located around 160~km below the photosphere \citep{Kum99}. For a 1~kG magnetic field strength the position of the $z_{eq}$ is around 80~km above the photosphere. Continuum intensity observations, as we have shown, are performed below this point, and therefore mode transmission/conversion at the VIRGO observational height has not yet occurred. This furthermore explains also why in the second and third $p$-band of the intensity observations we found an almost null size of suppression.
\subsection{Leaking fast and slow MHD waves to magnetic features}
We showed that the strength of the suppression in $p$-mode velocity amplitudes is highest when the solar activity is highest, which also holds for the strength of the acoustic enhancement in the HIPs bands, which is highest when the solar activity is highest. Therefore, both $p$-mode power suppression and HIPs power enhancement are correlated with the level of solar activity. We have learnt that mode conversion occurs in magnetic field strengths greater than $\approx 1$~kG and  that at a high ($ > 30^\circ$) field inclination the transmission coefficient from internal acoustic to slow magnetoacoustic becomes much less significant. This is extremely important to keep in mind if we want to link slow and fast MHD waves to the different solar magnetized areas. For example, magnetic regions have already been found to be strong absorbers of acoustic power in the frequency interval 2~mHz $<\nu<$ 4~mHz, and the suppression increases with the field strength \citep{Jai02}. Instead high-frequency waves seem to show up with a different behavior according to the magnetic field strength: the power is still suppressed in regions with strong magnetic fields, while areas surrounding them are found to be strong enhancers of acoustic emissivity above 5~mHz \citep{Hin98,Jai02,Nag07}. Why are regions of strong enhancement found around sunspots? Intense magnetic fields ($\approx$2~kG) with small field inclinations are generally confined to the umbra, and this physical condition meets the conditions for a larger transmission coefficient. The resonant modes (mainly between 2~mHz $<\nu<$ 4~mHz) are the ones that could undergo slow-mode transmission, because the $z_{eq}$ lies below their upper turning point, while in the same magnetic field strength the $z_{eq}$ for the traveling waves is below their excitation sources. As we go away from the umbra, the magnetic field strength decreases and as a consequence the magnetic field inclination becomes steeper \citep{Sol93}. These two conditions cause a decrease of the transmission coefficient, because by decreasing the magnetic field strength the $z_{eq}$ moves above their upper turning points. The peak transmission coefficient was actually found to occur close to a magnetic field inclination of $30^\circ$ for all frequencies \citep{Cro03}. But at the same time, the decreasing magnetic field strength moves the $z_{eq}$ above the excitation sources of traveling waves and
the increasing magnetic field inclination enhances the conversion from acoustic to fast-MHD waves. As a result the fast-MHD waves can be localized in the outer region of the sunspots, although recently observations have shown evidences of oscillations of 3 minutes inside sunspots \citep{Cen09}. Slow-MHD waves are mainly inside the sunspots.
  This mechanism can of course also explain the distribution of oscillations in and/or around active regions \citep{Jai02}. Recently, oscillations with frequencies between 2~mHz $<\nu<$ 8~mHz were found concentrated in bright plage areas \citep{Dew09,Osh09}.

Now that we have found a possible link between the solar magnetic features and slow/fast MHD waves, the results from integrated sunlight observations can help us to quantify how much of the solar surface is affected by mode conversion/transmission. Indeed, the strength of the low-frequency acoustic absorption and high-frequency acoustic enhancement is a direct measure of the total contribution coming from different magnetized areas of the solar surface. It has been shown that the solar surface coverage area of sunspots and plages over solar cycle 23 has been of 1$\%$ \citep{Alm04}. Therefore it will be extremely important to look for the missing contribution in the network and the intranetwork (Quiet Sun) areas \citep{Vec09}.
The understanding of the origin of low-frequency acoustic absorption and high-frequency acoustic enhancement is of particular interest nowadays, due to the possibility to use the induced solar cycle changes on $p$-mode parameters as precursor of solar activity. Acoustic waves seem indeed to undergo substantial changes before the appearance of macroscopic structure at the solar surface \citep{Cha07,How09,Sal09}. Therefore to correctly use the $p$-mode solar cycle changes as a precursor of activity it is extremely important to understand their origin and link them to the specific magnetized areas.

\section{Conclusions}
Low-frequency acoustic absorption in sunspots is very well known in the literature and mode conversion is nowadays believed to be the likely mechanism behind it. Localized sources of excitation have been proposed as a further mechanism to explain the observed high-frequency acoustic enhancement in and/or around sunspots. We decided to address this point by using 11 years of velocity and intensity integrated sunlight observations provided by BiSON and VIRGO and 6 years provided by the GOLF instrument, in an effort to search for changes to mode velocity amplitudes over the ascending phase of solar cycle 23. The results are dominated by the magnetism of the quiet Sun and only if low-frequency acoustic absorption and high-frequency acoustic enhancement occur in a significant area of the solar surface, we can expect to see evidence for both phenomen.

In agreement with previous findings, our results confirm a decrease of about 13$\%$ for resonant modes in the range of 2.5~mHz $<\nu<$ 3.5~mHz. With increasing frequency (3.5~mHz $<\nu<$ 6.8~mHz) there is a strong indication of a change of sign in the velocity amplitude suppression. We found in velocity observations at high-frequency (6.3~mHz $<\nu<$ 6.8~mHz) an acoustic enhancement of about 36$\pm$7 per cent, while in intensity it was about 5$\pm$2. This is the first time that integrated sunlight measurements have been characterized by the frequency dependence of velocity and intensity amplitudes on the solar cycle in the HIPs band. Therefore the novelty of this investigation  is the simultaneous observation of low-frequency acoustic absorption and high-frequency enhancement. Because of this finding, we can conclude that the same physical mechanism is behind in the Quiet Sun. The frequency dependence of the solar cycle changes in velocity amplitudes allowed us to compare the observations with the theoretical predictions of mode conversion. This comparison seems to point to mode conversion as the likely candidate to explain our findings. This statement could be strengthened by performing intensity observations at different  heights in the atmosphere, in order to check upon the nature of the waves. If we do not observe any high-frequency acoustic enhancement with height, this will imply that they are incompressible waves \citep{Cal08}, and we can rule out mode conversion as the physical mechanism behind low-frequency acoustic absorption and high-frequency acoustic enhancement. The wave behavior with height is even more extremely important to investigate another interesting issue. While fast magnetoacoustic waves are refracted backwards, slow-magnetoacoustic waves can travel in the atmosphere. Thus the latter could play a major role in the chromospheric heating \citep{Jef06}. 

This opens up a new interesting scenario: these observations have shown 
that the magnetism of the Quiet Sun affects the nature of the waves with 
increasing activity, which could even imply that
the chromospheric heating varies with the solar cycle. 
Nowadays, the nature of the Quiet Sun magnetism and 
its link with solar activity as the dependence of chromospheric heating 
on the state of the solar activity cycle is hotly debated \citep{Cat99,Alm09}. To work out 
this it will be extremely important to check upon the different 
contributions coming from regions which have a 
strong relatively uniform magnetic field, e.g. sunspots and active 
regions, with that (if any) from regions in which the field direction is 
very variable, like the intranetwork. The results of this further work 
will help us settle the debate about the origin 
of the quiet Sun magnetism. Furthermore the investigation 
 will also be important to check the 
possibility of using the solar cycle changes induced in the $p$-mode 
parameters as precursors of solar activity. We hope to extend this analysis during the peculiar extended minimum of activity
cycle 23, in which some frequency shifts are not following the traditional activity indexes \citep{Broo09, Sal09}.

Additionally, the interpretation of the observational findings in terms of 
mode conversion helps us to understand why 
 low-frequency acoustic enhancement is essentially located around sunspots and active 
regions. We have shown that the magnetic field strength and inclination 
control the locations where mode transmission and/or conversion might 
occur.
Transmission from acoustic to slow MHD waves between 2~mHz $<\nu<$ 4~mHz is 
more likely to occur within regions of a strong 
magnetic field and small field inclination. These 
 conditions are mainly satisfied in the umbra area. 
Instead conversion from acoustic to fast MHD waves above 5.8~mHz is more 
likely to occur around sunspots and active regions, because it 
requires a steep field inclination and a magnetic 
field strength of the order of 1~kG. These two conditions are satisfied in 
the sunspot penumbra. 
  The analysis carried out in this investigation shows that integrated 
sunlight measurements can investigate the solar-cycle 
changes induced in the $p$ and HIPs band. This has a further implication 
on asteroseismology. The technique developed here can be used to extend 
our understanding of stellar activity \citep{Kar07,Kar09}. Recent missions 
like Kepler aim to track variable stars for several years, which will 
give us the chance to follow their magnetic activity cycle by using 
intensity observations at low and high frequency. The analysis of the $p$-mode 
amplitude behavior with stellar activity cycle will give us a measure of 
the strength of the star activity compared to that of the Sun.

\begin{acknowledgements}
This work has been supported by the Swiss National Funding 200020-120114, by the Spanish grant PENAyA2007-62650 and the CNES/GOLF grant at the {\bf SAp}-CEA/Saclay. SOHO is an international cooperation between ESA and NASA. This paper also utilizes data collected by the ground-based BiSON network. We thank the members of the BiSON team, both past and present, for their technical and analysis support.
\end{acknowledgements}
\bibliographystyle{aa}
\bibliography{biblio1.bib}

\begin{thebibliography}{77}
\expandafter\ifx\csname natexlab\endcsname\relax\def\natexlab#1{#1}\fi

\bibitem[{{Alfv\'en}(1947)}]{Alf47}
{Alfv\'en}, H. 1947, MNRAS, 107, 211

\bibitem[{{Balmforth}(1990)}]{Bal90}
{Balmforth}, N.~J. $\&$~{Gough}, D.~O. 1990, \apj, 362, 256

\bibitem[{{Bloomfield} {et~al.}(2007){Bloomfield}, {Lagg}, \&
  {Solanki}}]{Blo07}
{Bloomfield}, D.~S., {Lagg}, A., \& {Solanki}, S.~K. 2007, \apj, 671, 1005

\bibitem[{{Braun} {et~al.}(1992){Braun}, {Duvall}, {Labonte}, \&
  et~al.}]{Bra92}
{Braun}, D.~C., {Duvall}, T.~D.~J., {Labonte}, B.~J., \& et~al. 1992, \apj,
  391, 113

\bibitem[{{Broomhall} {et~al.}(2009){Broomhall}, {Chaplin}, {Elsworth},
  {Fletcher}, \& {New}}]{Broo09}
{Broomhall}, A.~M., {Chaplin}, W., {Elsworth}, Y., {Fletcher}, S., \& {New}, R.
  2009, \apj, 700L

\bibitem[{{Brown} {et~al.}(1992){Brown}, {Bogdan}, {Lites}, \&
  {Thomas}}]{Bro92}
{Brown}, T.~M., {Bogdan}, T.~J., {Lites}, B.~W., \& {Thomas}, J.~H. 1992, \apj,
  394, 65

\bibitem[{{Cally}(1995)}]{Cal95}
{Cally}, P.~S. 1995, \apj, 451, 372

\bibitem[{{Cally}(2003)}]{Cal03}
{Cally}, P.~S. 2003, ASPC, 305, 152

\bibitem[{{Cally} \& Goosens(2008)}]{Cal08}
{Cally}, P.~S. \& Goosens, M. 2008, Sol.Phys, 251, 251

\bibitem[{{Carlsson} \& {Stein}(1997)}]{Car97}
{Carlsson}, M. \& {Stein}, R.~F. 1997, \apj, 481, 500

\bibitem[{{Cattaneo}(1999)}]{Cat99}
{Cattaneo}, F. 1999, ApJ, 515, 39

\bibitem[{{Centeno} {et~al.}(2009){Centeno}, {Collados}, \& {Trujillo
  Bueno}}]{Cen09}
{Centeno}, R., {Collados}, M., \& {Trujillo Bueno}, J. 2009, ApJ, 692, 1211

\bibitem[{{Chaplin} {et~al.}(1995){Chaplin}, {Elsworth}, {Howe}, \&
  et~al.}]{Cha95}
{Chaplin}, W.~J., {Elsworth}, Y.~P., {Howe}, R., \& et~al. 1995, MNRAS, 168, 1

\bibitem[{{Chaplin} {et~al.}(2003){Chaplin}, {Elsworth}, {Isaak}, \&
  et~al.}]{Cha03}
{Chaplin}, W.~J., {Elsworth}, Y.~P., {Isaak}, G., \& et~al. 2003, ApJ, 582,
  L115

\bibitem[{{Chaplin} {et~al.}(2000){Chaplin}, {Elsworth}, {Isaak}, \&
  {Miller}}]{Cha00}
{Chaplin}, W.~J., {Elsworth}, Y.~P., {Isaak}, G.~R., \& {Miller}, B.~A. 2000,
  MNRAS, 313, 32

\bibitem[{{Chaplin} {et~al.}(2007){Chaplin}, {Elsworth}, {Miller}, {Verner}, \&
  {New}}]{Cha07}
{Chaplin}, W.~J., {Elsworth}, Y.~P., {Miller}, B.~A., {Verner}, G.~A., \&
  {New}, R. 2007, ApJ, 659, 1749

\bibitem[{{Crouch} \& {Cally}(2003)}]{Cro03}
{Crouch}, A.~D. \& {Cally}, P.~S. 2003, Sol.Phys., 214, 201

\bibitem[{{Crouch} {et~al.}(2005){Crouch}, {Cally}, {Charbonneau}, \&
  {Desjardins}}]{Cro05}
{Crouch}, A.~D., {Cally}, P.~S., {Charbonneau}, P., \& {Desjardins}, M. 2005,
  AGU, 23, 4

\bibitem[{{De Pontieu} {et~al.}(2004){De Pontieu}, {Hansteen}, {Rouppe van der
  oort}, {van Noort}, \& {Carlsson}}]{Dep04}
{De Pontieu}, B., {Hansteen}, V.~H., {Rouppe van der oort}, L., {van Noort},
  M., \& {Carlsson}, M. 2004, \apj, 655, 624

\bibitem[{{De Wijn} \& {McIntosh}(2009)}]{Dew09}
{De Wijn}, A.~G. \& {McIntosh}, S.~W. 2009, \apjl, 702, 168

\bibitem[{{Deubner}(1975)}]{Deu75}
{Deubner}, L. 1975, \aap, 44, 371

\bibitem[{{Elsworth} {et~al.}(1993){Elsworth}, {Howe}, {Isaak}, {McLeod}, \&
  {Miller}}]{Els93}
{Elsworth}, Y.~P., {Howe}, R., {Isaak}, G.~R., {McLeod}, C.~P., \& {Miller},
  B.~A. 1993, MNRAS, 265, 888

\bibitem[{{Foglizzo} {et~al.}(1998){Foglizzo}, {Garc{\'{\i}}a}, {Boumier}, \&
  et~al.}]{Fog98}
{Foglizzo}, T., {Garc{\'{\i}}a}, R., {Boumier}, P., \& et~al. 1998, \aap, 330,
  341

\bibitem[{{Fossum} \& {Carlsson}(2005)}]{Fos05}
{Fossum}, A. \& {Carlsson}, M. 2005, Nature, 435, 919

\bibitem[{{Fr\"ohlich} {et~al.}(1995){Fr\"ohlich}, {Romero}, {Roth}, \&
  et~al.}]{Fro95}
{Fr\"ohlich}, C., {Romero}, J., {Roth}, H., \& et~al. 1995, \solphys, 162, 101

\bibitem[{{Gabriel} {et~al.}(1995){Gabriel}, {Grec}, {Charra}, \&
  et~al.}]{Gab95}
{Gabriel}, A.~H., {Grec}, G., {Charra}, J., \& et~al. 1995, \solphys, 162, 61

\bibitem[{{Garc\'ia} {et~al.}(1999){Garc\'ia}, {Jefferies}, {Toner}, \&
  {Pall\'e}}]{Gar99}
{Garc\'ia}, R.~A., {Jefferies}, S.~M., {Toner}, C.~G., \& {Pall\'e}, P.~L.
  1999, \aap, 346, L61

\bibitem[{{Garc\'ia} {et~al.}(1998b){Garc\'ia}, {Pall\'e}, {Turck-Chi\'eze}, \&
  et~al.}]{Gar98b}
{Garc\'ia}, R.~A., {Pall\'e}, P.~L., {Turck-Chi\'eze}, S., \& et~al. 1998b,
  \apj, 504, L51

\bibitem[{{Garc\'ia} {et~al.}(1998a){Garc\'ia}, {Roca-Cort\'es}, \&
  {R\'egulo}}]{Gar98a}
{Garc\'ia}, R.~A., {Roca-Cort\'es}, T., \& {R\'egulo}, C. 1998a, \aaps, 128,
  289

\bibitem[{{Garc\'ia} {et~al.}(2005){Garc\'ia}, {Turck-Chi\'eze}, {Boumier}, \&
  et~al.}]{Gar05}
{Garc\'ia}, R.~A., {Turck-Chi\'eze}, S., {Boumier}, P., \& et~al. 2005, \aap,
  442, 385

\bibitem[{{Goldreich} \& {Keeley}(1977)}]{Gol77}
{Goldreich}, P. \& {Keeley}, D.~A. 1977, \apj, 402, 721

\bibitem[{{Henney} {et~al.}(1999){Henney}, {Ulrich}, {Bertello}, {Bogart},
  {Bush}, \& et~al.}]{Hen99}
{Henney}, C.~J., {Ulrich}, R.~K., {Bertello}, L., {et~al.} 1999, \aap, 627

\bibitem[{{Hindman} {et~al.}(1998){Hindman}, {Jain}, \& {Zweibel}}]{Hin98}
{Hindman}, B.~W., {Jain}, R., \& {Zweibel}, E.~G. 1998, ApJ, 476, 392

\bibitem[{{Howe} {et~al.}(2009){Howe}, {Christensen-Dalsgaard}, {Hill}, \&
  et~al.}]{How09}
{Howe}, R., {Christensen-Dalsgaard}, J., {Hill}, F., \& et~al. 2009, ApJ, 701,
  87

\bibitem[{{Jain} \& {Haber}(2002)}]{Jai02}
{Jain}, R. \& {Haber}, D. 2002, \aap, 387, 1092

\bibitem[{{Jefferies} {et~al.}(2006){Jefferies}, {McIntosh}, {Armstrong}, \&
  et~al.}]{Jef06}
{Jefferies}, S.~M., {McIntosh}, S.~W., {Armstrong}, J.~D., \& et~al. 2006,
  \apj, 648, 151

\bibitem[{{Jim\'enez}(2006)}]{Jima06}
{Jim\'enez}, A. 2006, \apj, 646, 1398

\bibitem[{{Jim\'enez} {et~al.}(2005){Jim\'enez}, {Jim\'enez-Reyes}, \&
  {Garc\'ia}}]{Jima05}
{Jim\'enez}, A., {Jim\'enez-Reyes}, S.~J., \& {Garc\'ia}, R.~A. 2005, \apj,
  623, 1215

\bibitem[{{Jim\'enez-Reyes} {et~al.}(2007){Jim\'enez-Reyes}, {Chaplin}, \&
  {Elsworth}}]{Jim07}
{Jim\'enez-Reyes}, S.~J., {Chaplin}, W.~J., \& {Elsworth}, Y.~P. 2007, \apj,
  654, 1135

\bibitem[{{Jim\'enez-Reyes} {et~al.}(2004){Jim\'enez-Reyes}, {Chaplin},
  {Elsworth}, \& {Garc\'ia, R.~A.}}]{Jim04}
{Jim\'enez-Reyes}, S.~J., {Chaplin}, W.~J., {Elsworth}, Y.~P., \& {Garc\'ia,
  R.~A.} 2004, \apj, 604, 969

\bibitem[{{Jim\'enez-Reyes} {et~al.}(2003){Jim\'enez-Reyes}, {Garc\'ia},
  {Jim\'enez}, \& {Chaplin}}]{Jim03}
{Jim\'enez-Reyes}, S.~J., {Garc\'ia}, R.~A., {Jim\'enez}, A., \& {Chaplin},
  W.~J. 2003, \apj, 595, 446

\bibitem[{{Kalkofen}(2006)}]{Kal07}
{Kalkofen}, W. 2006, \apj, 671, 2154

\bibitem[{{Karoff}(2007)}]{Kar07}
{Karoff}, C. 2007, MNRAS, 381, 1001

\bibitem[{{Karoff} {et~al.}(2009){Karoff}, {Metcalfe}, {Chaplin}, \&
  et~al.}]{Kar09}
{Karoff}, C., {Metcalfe}, T.~S., {Chaplin}, W.~J., \& et~al. 2009, MNRAS, to be
  pubblished

\bibitem[{{Komm} {et~al.}(2000){Komm}, {Howe}, \& {Hill}}]{Kom00}
{Komm}, R.~W., {Howe}, R., \& {Hill}, F. 2000, \apj, 543, 472

\bibitem[{{Kosovichev} \& {Zharkova}(1998)}]{Kos98}
{Kosovichev}, A.~G. \& {Zharkova}, V.~V. 1998, Nature, 393, 317

\bibitem[{{Kumar} \& {Basu}(1999)}]{Kum99}
{Kumar}, P. \& {Basu}, S. 1999, \apj, 519, 396

\bibitem[{{Kumar} \& {Lu}(1991)}]{kum91}
{Kumar}, P. \& {Lu}, E. 1991, \apj, 375, L35

\bibitem[{{Kumar}(1990)}]{kum90}
{Kumar}, P. e.~a. 1990, in Lecture Notes in Physics, 367, 87

\bibitem[{{Lefebvre} {et~al.}(2008){Lefebvre}, {Garc\'ia}, {Jim\'enez-Reyes},
  {Turck-Chi\'eze}, \& {Mathur}}]{Lef08}
{Lefebvre}, S., {Garc\'ia}, R.~A., {Jim\'enez-Reyes}, S.~J., {Turck-Chi\'eze},
  S., \& {Mathur}, S. 2008, A$\&$A, 490, 1143

\bibitem[{{Leibacher} \& {Stein}(1971)}]{Lei71}
{Leibacher}, J.~W. \& {Stein}, R.~F. 1971, ApJL, 7, 191

\bibitem[{{Leighton} {et~al.}(1962){Leighton}, {Noyes}, \& {Simon}}]{Lei62}
{Leighton}, R.~B., {Noyes}, R.~W., \& {Simon}, G.~W. 1962, \apj, 135, 474

\bibitem[{{Lites} {et~al.}(1982){Lites}, {Chipman}, \& {White}}]{Lit82}
{Lites}, B.~W., {Chipman}, E.~G., \& {White}, O.~R. 1982, \apj, 253, 367

\bibitem[{{Nagashima}(2007)}]{Nag07}
{Nagashima}, Y.~Q. 2007, PASJ, 59, 631

\bibitem[{{O'Shea} \& {Doyle}(2009)}]{Osh09}
{O'Shea}, E. \& {Doyle}, G. 2009, AA, 494, 355

\bibitem[{{Salabert} {et~al.}(2009){Salabert}, {Garc\'ia}, {Pall\'e}, \&
  {Jim\'enez-Reyes}}]{Sal09}
{Salabert}, D., {Garc\'ia}, R.~A., {Pall\'e}, P.~L., \& {Jim\'enez-Reyes},
  S.~J. 2009, AA, 504, 1S

\bibitem[{{Salabert} \& {Jim\'enez-Reyes}(2006)}]{Sal06}
{Salabert}, D. \& {Jim\'enez-Reyes}, S.~J. 2006, ApJ, 650, 451

\bibitem[{{S\'anchez-Almeida}(2004)}]{Alm04}
{S\'anchez-Almeida}, J. 2004, ASPCS, 325, 100

\bibitem[{{S\'anchez-Almeida}(2009)}]{Alm09}
{S\'anchez-Almeida}, J. 2009, ASSC, 320, 121

\bibitem[{{Schunker} \& {Cally}(2006)}]{sch06}
{Schunker}, H. \& {Cally}, P.~S. 2006, MNRAS, 372, 551

\bibitem[{{Schwarzschild}(1948)}]{Sch48}
{Schwarzschild}, M. 1948, \apj, 107, 1S

\bibitem[{{Simoniello} {et~al.}(2004){Simoniello}, {Chaplin}, {Elsworth},
  {Isaak}, \& {New}}]{Sim04}
{Simoniello}, R., {Chaplin}, W.~J., {Elsworth}, Y.~P., {Isaak}, G., \& {New},
  R. 2004, \apj, 616, 594

\bibitem[{{Simoniello} {et~al.}(2009{\natexlab{a}}){Simoniello}, {Finsterle},
  {Garc\'ia}, \& {Elsworth}}]{Sim09a}
{Simoniello}, R., {Finsterle}, W., {Garc\'ia}, R.~A., \& {Elsworth}, Y.~P.
  2009{\natexlab{a}}, in: Synergies between Solar and Stellar modelling, Proc.
  HELAS Asteroseismology Workshop, Rome, Italy, in press

\bibitem[{{Simoniello} {et~al.}(2009{\natexlab{b}}){Simoniello}, {Finsterle},
  {Garc\'ia}, {Salabert}, \& {Jim\'enez}}]{Sim09b}
{Simoniello}, R., {Finsterle}, W., {Garc\'ia}, R.~A., {Salabert}, D., \&
  {Jim\'enez}, A. 2009{\natexlab{b}}, AIPC, 1170, 566S

\bibitem[{{Simoniello} {et~al.}(2008){Simoniello}, {Jim\'enez-Reyes},
  {Garc\'ia}, \& {Pall\'e}}]{Sim08}
{Simoniello}, R., {Jim\'enez-Reyes}, S.~J., {Garc\'ia}, R.~A., \& {Pall\'e},
  P.~L. 2008, AN, 329, 494

\bibitem[{{Socas-Navarro} {et~al.}(1998){Socas-Navarro}, {Ruiz-Cobo}, \&
  {TRujillo-Bueno}}]{Soc98}
{Socas-Navarro}, H., {Ruiz-Cobo}, B., \& {TRujillo-Bueno}, J. 1998, ApJ, 507,
  470

\bibitem[{{Solanki} \& {Schmidt}(1993)}]{Sol93}
{Solanki}, S.~K. \& {Schmidt}, H.~U. 1993, ApJ, 267, 287

\bibitem[{{Stenflo}(1982)}]{ste82}
{Stenflo}, J.~O. 1982, \solphys, 80, 209

\bibitem[{{Thomas} \& {Stanchfield}(2000)}]{Tho00}
{Thomas}, J.~H. \& {Stanchfield}, D.~C.~H. 2000, \apj, 537, 1087

\bibitem[{{Turck-Chi\'eze} {et~al.}(2004){Turck-Chi\'eze}, {Garc\'ia},
  {Couvidat}, \& et~al.}]{Tur04}
{Turck-Chi\'eze}, S., {Garc\'ia}, R.~A., {Couvidat}, \& et~al. 2004, \apj, 604,
  455

\bibitem[{{Ulmschneider} \& {Narain}(1990)}]{Ulm90}
{Ulmschneider}, P. \& {Narain}, U. 1990, IAUS, 142, 97

\bibitem[{{Ulrich}(1970)}]{Ulr70}
{Ulrich}, R.~K. 1970, \apj, 162, 993

\bibitem[{{Ulrich} {et~al.}(2000){Ulrich}, {Garc\'ia}, {Robillot}, \&
  et~al.}]{Ulr00}
{Ulrich}, R.~K., {Garc\'ia}, R.~A., {Robillot}, J.~M., \& et~al. 2000, \aap,
  364, 799

\bibitem[{{Unno}(1959)}]{Unn59}
{Unno}, W. 1959, \apj, 129, 375

\bibitem[{{Vecchio} {et~al.}(2009){Vecchio}, {Cauzzi}, \& {Reardon}}]{Vec09}
{Vecchio}, A., {Cauzzi}, G., \& {Reardon}, K. 2009, A$\&$A, 494, 629

\bibitem[{{Viereck} {et~al.}(2001){Viereck}, {McMullin}, {Darrell}, {Weber}, \&
  {Tobiska}}]{Ver01}
{Viereck}, R., {McMullin}, D., {Darrell}, J., {Weber}, M., \& {Tobiska}, W.~K.
  2001, Geo RL, 28, 1343

\bibitem[{{Woods} \& {Cram}(1981)}]{Woo81}
{Woods}, D.~T. \& {Cram}, L.~E. 1981, \solphys, 69, 233

\end{thebibliography}


18 gid=2048330233
17 uid=174949570
20 ctime=1271750678
20 atime=1271750680
38 LIBARCHIVE.creationtime=1271663599
24 SCHILY.dev=234881026
22 SCHILY.ino=2606405
18 SCHILY.nlink=1

\end{document}